\documentclass[a4paper]{article}
\pdfoutput=1
\usepackage{etex} 
\reserveinserts{28}
\usepackage{pdflscape}

\usepackage{url}

\usepackage{xcolor}
\definecolor{amazing}{RGB}{254,67,101}

\usepackage{bbm}
\usepackage{mathrsfs}
\usepackage{latexsym}
\usepackage{paralist}
\usepackage{microtype}

\usepackage{suffix}

\usepackage{xpunctuate}
\usepackage[all,british]{foreign} 
\redefnotforeign[ie]{i.e\xperiodafter} 
\redefnotforeign[eg]{e.g\xperiodafter}

\usepackage{xfrac} 
\usepackage[style=english,english=british]{csquotes}

\usepackage{amsmath, amsthm, amssymb, amsfonts} 
\usepackage{thmtools, thm-restate}
\usepackage{mathtools}

\usepackage[full,small]{complexity}

\usepackage[notref,color,final]{showkeys} 

\usepackage{xargs}
\usepackage{afterpage} 

\PassOptionsToPackage{dvipsnames,usenames,table}{xcolor}
\usepackage{tikz}
\usetikzlibrary{calc}

\usepackage{booktabs} 
\usepackage{longtable}
\usepackage{float}
\usepackage{wrapfig}
\usepackage{floatflt}
\usepackage{framed}
\usepackage{ctable}
\usepackage{dcolumn}
\usepackage{adjustbox}

\usepackage[shortlabels]{enumitem} 
\usepackage{xspace}
\usepackage{xifthen}
\usepackage{fixltx2e}
\usepackage{url}
\usepackage{scrtime}

\usepackage[longend,vlined]{algorithm2e}

\usepackage{xkvltxp}
\usepackage[draft,english,silent]{fixme}
\newcommand{\todo}[1]{\fxfatal{\color{red}#1}}

\usepackage{index}

\makeindex
\newindex{sym}{sym.idx}{sym.ind}{Symbol index}
\newindex{prob}{prob.idx}{prob.ind}{Problem index}

\newcommand\mathindex[1]{\index[sym]{\ensuremath{{#1}}}}
\WithSuffix\newcommand\mathindex*[1]{\index*[sym]{\ensuremath{{#1}}}}

\makeatletter
\newcommand{\raisemath}[1]{\mathpalette{\raisem@th{#1}}}
\newcommand{\raisem@th}[3]{\raisebox{#1}{$#2#3$}}
\makeatother

\def\N{\mathbb{N}} 

\newclass{\paraNP}{paraNP}

\newcommand\restr[2]{{
  \left.\kern-\nulldelimiterspace 
  #1 
  \vphantom{\big|} 
  \right|_{#2} 
  }}

\def\isom{\simeq}
\def\subgraph{\subseteq}

\def\sminor^#1{
    \preccurlyeq_{\mathrlap{\mathsf{m}}}^{#1}
} 
\def\ssminor^#1{%
    \mathbin{\dot\preccurlyeq_{{\mathrlap{\mathsf{m}}}}^{#1}}\,
} 
\def\stminor^#1{
    \preccurlyeq_{\mathrlap{\mathsf{t}}}^{#1}
} 
\def\sstminor^#1{%
    \mathbin{\dot\preccurlyeq_{{\mathrlap{\mathsf{t}}}}^{#1}}\,
} 

\def\dir#1{\vec{#1}}
\def\grad_#1{\nabla\!_{#1}}
\def\sgrad_#1{{\dot\nabla}\!_{#1}}
\def\topgrad_#1{\widetilde \nabla\!_{#1}}
\def\stopgrad_#1{\dot{\widetilde\nabla}\!_{#1}}
\def\topomega_#1{\widetilde \omega_{#1}}

\def\colnum_#1{ \operatorname{col}_{#1} }
\def\wcolnum_#1{ \operatorname{wcol}_{#1} }
\def\adm_#1{ \operatorname{adm}_{#1} }

\renewcommand{\leq}{\leqslant}

\renewcommand{\geq}{\geqslant}

\renewcommand{\epsilon}{\varepsilon}

\renewcommand{\emptyset}{\varnothing}

\newlength{\convarrowwidth}
\usepackage{stmaryrd}

\newcommand{\widthm}[1]{ \mathbf{#1} } 

\DeclareMathOperator{\width}{ \widthm{width} } 

\def\YYYY{{Y_0 \uplus Y_1 \uplus \cdots \uplus Y_\ell}} 
\WithSuffix\def\YYYY'{{Y'_0 \uplus Y'_1 \uplus \cdots \uplus Y'_{\ell'}}}

\usepackage{pifont}
\newcommand{\yaay}{\kern4pt \ding{51} \kern-8pt \ding{51}}%

\def\any{\mathord{\color{black!33}\bullet}}%
\def\root{\operatorname{root}} 
\def\rpath{\operatorname{rpath}} 
\def\leaves{\operatorname{leaves}} 
\def\depth{\operatorname{depth}}
\def\height{\operatorname{height}}
\def\anc{\operatorname{anc}}

\def\conv{ \mathbin{ \ast } } 

\theoremstyle{plain}
\newtheorem{lemma}{Lemma}

\newtheorem{theorem}{Theorem}
\newtheorem{corollary}{Corollary}
\newtheorem{observation}{Observation}
\newtheorem{proposition}{Proposition}
\newtheorem*{claim}{Claim}

\newtheoremstyle{case}
  {\topsep}   
  {\topsep}   
  {}  
  {\parindent}       
  {\bfseries} 
  {\normalfont.}         
  {5pt plus 1pt minus 1pt} 
  {#1 #2: {\normalfont #3}}          

\theoremstyle{case}

\numberwithin{subcase}{case}
\makeatletter
\@addtoreset{case}{lemma}
\@addtoreset{case}{theorem}
\@addtoreset{case}{corollary}
\makeatother

\theoremstyle{definition}

\newtheorem{definition}{Definition}

\newcommand*\varrule[1][0.4pt]{\leavevmode\leaders\hrule height#1\hfill\kern0pt}

\setlist[1]{labelindent=\parindent,leftmargin=*} 
\setlist{itemsep=0pt}
\setitemize[1]{label={\small\textbullet}}

\newenvironment{tightcenter}
 {\parskip=0pt\par\nopagebreak\centering}
 {\par\noindent\ignorespacesafterend}

\usepackage{ctable}
\newlength{\RoundedBoxWidth}
\newsavebox{\GrayRoundedBox}
\newenvironment{GrayBox}[1]%
   {\setlength{\RoundedBoxWidth}{\textwidth-4.5ex}
    \def\boxheading{#1}
    \begin{lrbox}{\GrayRoundedBox}
       \begin{minipage}{\RoundedBoxWidth}%
   }{%
       \end{minipage}
    \end{lrbox}%
    \begin{tightcenter}%
    \begin{tikzpicture}%
       \node(Text)[draw=black!20,fill=white,rounded corners,%
             inner sep=2ex,text width=\RoundedBoxWidth]%
             {\usebox{\GrayRoundedBox}};
        \coordinate(x) at (current bounding box.north west);
        \node [draw=white,rectangle,inner sep=3pt,anchor=north west,fill=white] 
        at ($(x)+(6pt,.75em)$) {\boxheading};
    \end{tikzpicture}
    \end{tightcenter}\vspace{0pt}%
    \ignorespacesafterend
}

\newenvironment{problem*}[2][]{\noindent\ignorespaces%
                                \FrameSep=6pt%
                                \parindent=0pt%
                \vspace*{-.5em}
                \ifthenelse{\isempty{#1}}{%
                  \begin{GrayBox}{\textsc{#2}}%
                }{%
                  \begin{GrayBox}{\textsc{#2} parametrised by~{#1}}%
                }
                \index[prob]{#2@\textsc{#2}|textbf}%
                \begin{tabular*}{\textwidth}{@{\hspace{.1em}} >{\itshape} p{1.6cm} p{0.8\textwidth} @{}}%
            }{
                \end{tabular*}%
                \end{GrayBox}%
                \vspace*{-.5em}
                \ignorespacesafterend
            }       

\renewenvironmentx{leftbar}[2][1=0.5pt, 2=5pt]{%
  \def\FrameCommand{\vrule width #1 \hspace{#2}}\MakeFramed {\advance\hsize-\width \FrameRestore}}%
{\endMakeFramed}

\newcommand*\circled[4]{\tikz[baseline=(char.base)]{
             \node[circle,fill=#3,draw=#4,draw,inner sep=1pt,minimum size=1.2em] (char) {\color{#2} #1};}}

\newlength{\wleft}  \newlength{\wright}

\definecolor{Maroon}{cmyk}{0, 0.87, 0.68, 0.32}
\definecolor{RoyalBlue}{cmyk}{1, 0.50, 0, 0}
\definecolor{Black}{cmyk}{0, 0, 0, 0}
\definecolor{White}{rgb}{1, 1, 1}

\makeatletter
\let\orgdescriptionlabel\descriptionlabel
\def\@savelabel{}
\renewcommand*{\descriptionlabel}[1]{%
  \let\orglabel\label
  \let\label\@gobble
  \phantomsection
  \def\@savelabel{#1}
  \edef\@currentlabel{{\def\hfil{}#1}}
  \edef\@currentlabelname{#1}%
  \let\label\orglabel
  \orgdescriptionlabel{#1}%
}
\makeatother
\makeatletter
\def\namedlabel#1#2{\begingroup
   \def\@currentlabel{#1}%
   \label{#2}\endgroup
}
\makeatother

\usepackage{hyphenat}
\renewcommand{\th}{%
    \ifmmode
        ^\mathrm{th}%
    \else%
        \textsuperscript{th}\xspace%
    \fi%
}
\newcommand{\st}{%
    \ifmmode
        ^\mathrm{st}%
    \else%
        \textsuperscript{st}\xspace%
    \fi%
}
\newcommand{\nd}{%
    \ifmmode
        ^\mathrm{nd}%
    \else%
        \textsuperscript{nd}\xspace%
    \fi%
}
\newcommand{\rd}{%
    \ifmmode
        ^\mathrm{rd}%
    \else%
        \textsuperscript{rd}\xspace%
    \fi%
}

\def\Nesetril{Ne\v{s}et\v{r}il\xspace}

\def\Dvorak{Dvo\v{r}\'{a}k\xspace}

\def\Tuma{T{\r{u}}ma\xspace}

\usepackage{environ}%
\NewEnviron{nofiitable}{\noindent\ignorespaces%
  
  \[\arraycolsep=1.4pt%
  \begin{array}{rcr p{1cm} rcr}
    \BODY
  \end{array}
  \]
}

\newcolumntype{m}{>{$}l<{$}}
\newcolumntype{M}{>{$\displaystyle}l<{$}} 

\newcolumntype{L}{l}  
\newcolumntype{C}{c}  
\newcolumntype{R}{r}  
\newcolumntype{X}{>{\global\let\currentrowstyle\relax}}
\newcolumntype{^}{>{\currentrowstyle}}

\usepackage{authblk}

\title{A color-avoiding approach to subgraph counting in bounded expansion classes}
\author[1]{Felix~Reidl}
\author[2]{Blair~D.~Sullivan}
\affil[1]{\texttt{f.reidl@dcs.bbk.ac.uk} \protect\\ Birkbeck, University of London}
\affil[2]{\texttt{sullivan@cs.utah.edu} \protect\\ School of Computing, University of Utah}

\usepackage{centernot}

\def\order{\leq}
\def\porder{\preccurlyeq}
\def\sporder{\prec}

\let\anc\relax
\DeclareMathOperator{\anc}{\porder_\mathsf{anc}}
\DeclareMathOperator{\rel}{\mathsf{relax}}
\DeclareMathOperator{\stem}{\mathsf{stem}}
\DeclareMathOperator{\hint}{\mathsf{hint}}
\DeclareMathOperator{\defects}{\mathcal D}
\def\hash_#1{%
  \mathchoice{\mathop{\#}\limits_{#1}} 
             {\mathop{\#}_{#1}} 
             {\mathop{\#}_{#1}} 
             {\mathop{\#}_{#1}} 
}
\def\thash_#1{%
  \mathchoice{\mathop{\widehat\#}\limits_{#1}} 
             {\mathop{\widehat\#}_{#1}} 
             {\mathop{\widehat\#}_{#1}} 
             {\mathop{\widehat\#}_{#1}} 
}
\def\counting{\ensuremath{\mathsf \#}}

\DeclareMathOperator{\ET}{ET}

\def\subtog{\subseteq}

\def\G{\mathbf G}
\def\H{\mathbf H}
\def\K{\mathbf K}
\def\D{\mathbf D}

\def\GG{\mathbb G}
\def\HH{\mathbb H}
\def\KK{\mathbb K}
\def\DD{\mathbb D}

\newcommand{\CC}{\mathsf C}

\def\xtox{\bar x \mapsto \bar x}
\def\xtoy{\bar x \mapsto \bar y}

\def\embeds#1{\xhookrightarrow{\raisebox{-.8pt}[.8\height][0pt]{\scriptsize$\,#1\;$}}}
\def\notembeds{\not\hookrightarrow}

\def\commentmark#1{\circled{#1}{gray}{white}{gray}}
\def\comment#1#2{\textcolor{gray}{\commentmark{#1} #2}}

\let\oldmarginpar\marginpar
\renewcommand\marginpar[1]{\-\oldmarginpar[\raggedleft\footnotesize #1]%
{\raggedright\footnotesize #1}}

\begin{document}

\maketitle
\begin{abstract}
  \noindent

  We present an algorithm to count the number of occurrences
  of a pattern graph~$H$ as an induced subgraph in a host graph~$G$.
  If~$G$ belongs to a bounded expansion class, the algorithm runs in
  linear time. Our design choices are motivated by the need for
  an approach that can be engineered into a practical implementation
  for sparse host graphs.

  Specifically, we introduce a decomposition of the pattern~$H$ called
  a \emph{counting dag}~$\dir C(H)$ which encodes an order-aware,
  inclusion-exclusion counting method for~$H$. Given such a counting dag
  and a suitable linear ordering~$\GG$ of~$G$ as input, our algorithm
  can count the number of times~$H$ appears as an induced subgraph in~$G$
  in time $O(\|\dir C\| \cdot h  \wcolnum_{h}(\GG)^{h-1} |G|)$,
  where~$\wcolnum_h(\GG)$ denotes the maximum size of the
  weakly $h$-reachable sets in~$\GG$. This implies, combined with previous results, an algorithm
  with running time~$O(4^{h^2}h (\wcolnum_h(G)+1)^{h^3} |G|)$ which only
  takes~$H$ and~$G$ as input. \looseness-1

  We note that with a small modification, our algorithm can instead use strongly $h$-reachable sets
  with running time $O(\|\dir C\| \cdot h  \colnum_{h}(\GG)^{h-1} |G|)$, resulting in
  an overall complexity of~$O(4^{h^2}h \colnum_h(G)^{h^2} |G|)$ when only given $H$ and $G$.

  Because orderings with small weakly/strongly reachable sets can be computed relatively efficiently in practice~\cite{WcolExperimental}, our algorithm provides a promising alternative to algorithms using the traditional $p$-treedepth colouring framework~\cite{TooManyColours}.  We describe preliminary experimental results from an initial open source implementation which highlight its potential.
\end{abstract}

\section{Introduction}

We consider the problem of counting the number of times a \emph{pattern}  graph
$H$ appears in a \emph{host} graph $G$ as an induced subgraph.
Without any restrictions on $G$, this problem is already difficult for very
simple $H$: Flum and Grohe~\cite{ParamCounting} showed that it is
\counting\W[1]-hard when $H$ is a clique and Chen and Flum showed that it is
\counting\W[2]-hard when it is a path~\cite{CountingPathHard}
(\counting\W[1]-hard if we drop the requirement of being an \emph{induced}
subgraph).  That is, there is little hope for algorithms with running time
$f(|H|) \cdot \poly{|G|}$ for these problems unless \eg counting satisfying
assignments of a 3-CNF formula is possible in time $2^{o(n)}$ (further details
on parameterized counting classes can be found in Flum and Grohe's
book~\cite{FlumGrohe}).

The situation is less glum when we restrict ourselves to sparse host graphs.
For example, Eppstein, L{\"{o}}ffler, and Strash showed that enumerating all
cliques in a $d$-degenerate host graph~$G$ is possible in time $O(d \cdot 3^{d/3}
|G|)$~\cite{EppsteinDegenCliques}. More generally, we can count  any pattern
graph $H$ on $h$ vertices in time $O(f(h) \cdot |G|)$ provided that $G$ is taken
from a graph class of \emph{bounded expansion} (where $f$ depends on the class)
and time $O(f(h) \cdot |G|^{1+o(1)})$ if it is taken from a \emph{nowhere dense}
graph class.

Currently, two types of approaches exist in these sparse settings.  One class of
algorithms is based on so-called \emph{$p$-treedepth colourings}: given a class
$\mathcal G$ of bounded expansion we can colour any $G \in \mathcal G$ in time
$f(p) \cdot |G|$ with $f(p)$ colours so  that any subgraph of $G$ with $i < p$
colours has treedepth $\leq i$. By computing an $h$-treedepth colouring this
effectively reduces the problem to counting $H$ in a graph~$G'$ of treedepth $t
\leq |H|$. Ossona de Mendez and \Nesetril, who also introduced the notion of
bounded expansion and nowhere dense classes, presented an algorithm for this
latter step with a running time of $O(2^{ht}ht \cdot |G'|)$ ~\cite{Sparsity};
with Demaine, Rossmanith, S{\'a}nchez Villaamil, and Sikdar we later improved
this to $O(6^h t^h h^2 \cdot |G'|)$~\cite{SparsityNetworks}.  Using this
subroutine, we can count occurrences of $H$ in $G$ by first computing an
$h$-treedepth colouring with $f'(h)$ colours, then iterate through all
$\sum_{i=1}^h {f'(h) \choose i}$ colour combinations and count in time $O(6^h
t^h h^2 \cdot |G'|)$ the number of times $H$ appears in the subgraph $G'$
induced by these colours. The final count is then computed via
inclusion-exclusion over the counts obtained for the colour sets.

While conceptually simple, it turns out that these algorithms are currently
impractical: a) computing $h$-treedepth colourings is currently computationally
quite expensive and b) the number of colours $f'(h)$ is so big that already the
act of enumerating all relevant colour subsets takes too
long~\cite{TooManyColours}.  It turns out that the underlying technique for
these algorithms---so-called transitive-fraternal augmentations~\cite{Sparsity}
(tf-augmentations) with some practical and
improvements~\cite{FelixThesis,TooManyColours}---also lies at the heart of the
other available technique.  Kazana and Segoufin used tf-augmentations to
enumerate first-order queries with constant delay (or to count such queries in
linear time) in classes with bounded expansion~\cite{FOEnumBndExp} and \Dvorak
and \Tuma designed a dynamic data structure\footnote{To be precise this data
structure only uses fraternal augmentations.} to count subgraphs with amortized
polylogarithmic updates~\cite{SubgraphDynamic}.  The latter approach also has
the drawback that in order to count induced subgraphs, one must perform a big
inclusion-exclusion over all supergraphs of the pattern.

Despite our best efforts to make tf-augmentations \emph{practical}, so far they
seem to be only useful in very tame settings like bounded-degree
graphs~\cite{SGC}.  It is thus natural to ask whether we can solve the
subgraph-counting problem \emph{without} relying on $p$-treedepth colourings or
even tf-augmentations. In particular, the computation of so-called
\emph{generalized colouring numbers} (a set of graph measures
introduced by Kierstead and Yang~\cite{WCol}
which provide an
alternative characterisation of bounded expansion/nowhere dense
classes~\cite{WColBndExp}), appears much more feasible in
practice~\cite{WcolExperimental}, and offers an attractive ordering-based alternative.

Our contribution here is to provide an algorithm to count induced subgraphs
which is solely based on the \emph{weak colouring number} (or the
\emph{colouring number}). At a high level, we do this by using a suitable linear
order of the host graph and counting how often each of the possible pattern
graph orders appears in it\footnote{We view these orderings as a type of graph
decomposition and therefore assume they are part of the input.}.  The crucial
insight here is that under some orderings, the pattern graph can only appear
inside certain neighbourhood-subsets and that all other orderings can be reduced
to these easily countable cases via inclusion-exclusion style arguments. Note
that in contrast to \Dvorak and \Tuma's approach, the objects in our
inclusion-exclusion are specific ordered graphs and we can therefore avoid
counting \emph{all} supergraphs of the pattern.

In order to establish the practicality of our approach, we implemented a
prototype of the entire algorithmic pipeline described in this paper using a
combination of Rust and Python.  The code is available under a BSD 3-clause
license at \url{http://www.github.com/theoryinpractice/mandoline}.

We begin in Section~\ref{sec:prelim} by providing necessary definitions and notation related to
ordered graphs, reachability and bounded expansion. We then describe our
approach to decomposing the pattern graph and combining counts of partial matches
in Section~\ref{sec:patterns}. We combine these subroutines with a new data
structure in Section~\ref{sec:algorithms} to form the basis of our linear-fpt
algorithm. Finally, in Section~\ref{sec:discussion}, we briefly discuss our
experimental results and future work.

\section{Preliminaries}\label{sec:prelim}

\subsubsection*{Trees}

All trees in this paper will be assumed to be rooted. In particular, a
\emph{subtree} is always a rooted subtree. For a tree~$T$, we write $\root(T)$
to denote its root and $\leaves(T)$ to denote its leaves. The \emph{root path}
$\rpath_T(x)$ for a node $x \in T$ is the unique path from $\root(T)$ to $x$
in $T$.

The \emph{ancestor relationship} $\anc^{\kern-10ptT}\kern4pt$ of a tree~$T$ is
the partial order defined via
\[
  x \anc^{\kern-10ptT}\kern4pt y \iff x \in \rpath_T(y).
\]

\subsubsection*{Partial and total orders}

\marginpar{Digraph representation}
partial orders and the symbol $\sporder$ to denote the relation $(x \porder y)
\land (x \neq y)$.  Given a partial order $\porder$ over $S$, its \emph{digraph
representation} is a dag with vertices $S$ and arcs
$\{ xy \in S \times S \mid x \sporder y\}$.

\marginpar{Principal digraph}
The \emph{principal digraph} of a partial order~$\porder$ over $S$ is the
dag with vertices $S$ and the arcs
\[
  \{ xy \in S \times S \mid x \sporder y
      ~\text{and there is no $z \in S$ with}~x \sporder z \sporder y\}
\]
Note that if $\dir D$ is the principal digraph of $\porder$, $S$; then the
transitive closure of $\dir D$ is the digraph representation of $\porder$, $S$.

\marginpar{Tree (order)} A partial order $\porder$ over $S$ is a \emph{tree} if
for every element $x \in S$, the set $\{ y \mid y \porder x \}$ is well-ordered
by $\porder$. Alternatively, $\porder$ is a tree if its principal digraph its a
directed tree, \eg all arcs are oriented away from the root node.

\marginpar{Linear extension}
A \emph{linear extension} of $\porder$ is a total order $\leq$ such that
$x \porder y$ implies $x \leq y$. The linear extensions of $\porder$ are precisely
the topological orderings of either is digraph representation or its principal
digraph.

\subsubsection*{Ordered graphs}

A \emph{tree ordered graph} $\G = (G, \porder)$ (tog) is a graph whose vertex set
$V(\G) := V(G)$ is imbued with a (partial) order relation $\porder$ with the following
properties:
\begin{enumerate}
  \item The relation $\porder$ is a \emph{tree order}.
  \item The relation $E(G)$ is \emph{guarded} by $\porder$: for every edge
  $uv \in E(G)$ it holds that either $u \porder v$ or $v \porder u$.
\end{enumerate}
We define~$T(\G)$ to be the tree-representation of $\porder$ with node
set $V(\G)$. We extend the notions and notation of roots, leaves, and root-paths
to togs via $\root(\G) := \root(T(\G))$, $\leaves(\G) := \leaves(T(\G))$, and
$\rpath_\G(\any) = \rpath_{T(\G)}(\any)$. Given a tog $\G$ we write $\porder_\G$
to denote its tree-order relation and we will use the notation $u \sporder_\G v$
to mean that $u \porder_\G v$ and $u \neq v$.
\marginpar{Ordered vertex set}
An \emph{ordered vertex set} $\bar x := x_1,\ldots,x_\ell$ of a tog $\G$ is
a sequence of vertices which satisfies $x_1 \sporder_\G x_2 \sporder_\G \ldots
\sporder_\G x_\ell$. The \emph{length} of an ordered vertex set $\bar x$ is the
number of elements in it. We use the symbol $\emptyset$ to denote both the empty
set and the empty ordered vertex set and make sure it is clear from the context which
is meant.

\marginpar{Linear graph}
If $\porder_\G$ is a total order we call $\G$ a \emph{linear graph}.  We
will use the symbol $\GG$ (instead of $\G$) and $\order_{\GG}$ (instead of $\porder_{\G}$)
in cases were we want to emphasize that the ordering is linear. For a given
graph $G$ we write $\Pi(G)$ for the set of all linear graphs obtained from $G$
by permuting its vertex set.

\marginpar{Subtog}
A tog isomorphism $\H \isom \G$ is a bijection between the vertex sets of $\H$
and $\G$ that preserves both the edge and the ordering relations.
Given a vertex set $X \subseteq V(\G)$,
the tog induced by~$X$, denoted by $\G[X]$, is the tog
$(G[X], \porder_\G\!|_X)$. In general, a tog $\H$ is an \emph{induced
subtog} of a tog $\G$ if there
exists a vertex set $X$ such that $\H \isom \G[X]$ and we
write $\H \subtog \G$.

\marginpar{Stem}
The \emph{stem} of a tog $\G$ is the ordered set $\bar x$ of maximal length
such that $\bar x$ is linearly ordered under $\porder_\G$ and $\max \bar x
\porder_\G u$ for all vertices $u \in V(\G) - \bar x$. If we visualize
$\porder_\G$ as a tree then the stem is the path from the root to the
first node with more than one child.

\marginpar{embeds}
We say that a tog $\H$ \emph{embeds} into a tog $\G$ if
there exists a subgraph isomorphism~$\phi$ from $H$ to $G$ that further
satisfies
\[
  u \porder_\H v \implies \phi(u) \porder_\G \phi(v)
\]
\marginpar{$\embeds{\phi}$, $\embeds{}$}
and we write $\H \embeds{\phi} \G$ or $\H \embeds{} \G$ if we
do not need to assign a variable to the embedding.

\marginpar{$\min_\G$, $\max_\G$}
For a vertex set
$X \subseteq V(\G)$, we let $\min_\G X$ and $\max_\G X$
be the minimum and maximum according to $\porder_\G$
(if they exist). We extend this notation to subtogs via
$\min_\G \H := \min_\G V(\H)$ and $\max_\G \H := \max_\G V(\H)$.
Note that by the two properties of togs, every vertex set that induces a
connected subtog necessarily has a minimum. Moreover, such a minimum
is preserved by tog embeddings:

\begin{observation}\label{obs:min-embed}
	Let $\H \embeds{\phi} \G$ and let $\H' \subtog \H$ such that
	$\min_\H \H'$ exists. Then $\min_\G \phi(\H') = \phi(\min_\H \H')$.
\end{observation}
\begin{proof}
	From $\min_\H \H' \porder_\H u$ for every $u \in \H'$ and the fact
	that $\phi$ is an embedding we conclude that
	$\phi(\min_\H \H') \porder_\G \phi(u)$ for every $u \in \H'$.
	Therefore $\min_\G \phi(\H') = \phi(\min_\H \H')$.
\end{proof}

\noindent
This in particular implies that embeddings preserved ordered vertex sets:
if $\bar x$ is an ordered vertex set of $\H$ and $\H \embeds{\phi} \G$,
then $\phi(\bar x)$ is an ordered vertex set of $\G$.
Finally, we note that embeddings are transitive:

\begin{observation}\label{obs:trans-embed}
	If $\K \embeds{\phi} \H$ and $\H \embeds{\psi} \G$ then
	$\K \embeds{\psi \,\circ\, \phi} \G$.
\end{observation}

\noindent
The notion of \emph{elimination trees} (known also under the name \emph{treedepth
decomposition} and many others) connects tree ordered graphs to linearly
ordered graphs.

\begin{definition}[Elimination tree]
  Given a connected linearly ordered graph $\HH$, the \emph{elimination tree}
  $\ET(\HH)$ is defined recursively as follows: Let $x := \min \HH$ and
  let $\KK_1,\ldots,\KK_s$ be the connected components of $\HH - x$.
  Then $\ET(\HH)$ has $x$ as its root with the roots of $\ET(\KK_1),\ldots,
  \ET(\KK_s)$ as its children.
\end{definition}

\begin{definition}[Tree order relaxation]
  Given a connected linearly ordered graph $\HH$ and its elimination
  tree $T := \ET(\HH)$, we define its \emph{tree order relaxation} as the
  tog $\rel(\HH) = (H, \anc^{\kern-10ptT}\kern4pt)$.
\end{definition}

\noindent
Observe that $\rel(\HH) \embeds{~~} \HH$ and these embeddings
have the stem of $\rel(\HH)$ as fixed points.

\marginpar{etog}
\begin{definition}[Elimination-ordered graph (etog)]
  A tog $\H$ for which there exists a linear graph $\HH$ such
  that $\rel(\HH) = \H$ is called an \emph{elimination-ordered graph} (etog).
\end{definition}

\begin{lemma}\label{lemma:subtree-path}\marginpar{$\rel(\cdot)$}
  Let $\H = \rel(\HH)$ be a tree order relaxation of a connected
  linear graph $\HH$. Then for every pair of vertices $x,y \in V(\H)$ it
  holds that that $x \porder_\H y$ if and only if there exists an $x$-$y$-path $P$ with
  $\min_\H P = x$.
\end{lemma}
\begin{proof}
  Let $T = \ET(\HH)$ be the elimination tree whose ancestor relationship
  defines $\porder_\H$. First assume that $x \porder_\H y$.
  By definition, the nodes of the subtree~$T_x$ induce
  a connected subtog of $\HH$ and hence in $\H$, thus there exists a path $P$ from
  $x$ to $y$ in $\H[V(T_x)]$ and hence $\min_\H P = x$.

  Now assume that either $y \porder_\H x$ or that $x$ and $y$ are
  incomparable under $\porder_\H$. Assume towards a contradiction that
  there exists path $P$ in $\H$ with $\min_\H P = x$. But then
  $x \porder_\H y$ since $y \in P$, a contradiction.
\end{proof}

\begin{corollary}\label{cor:incomparable-path}
  Let $\H = \rel(\HH)$ be a tree order relaxation of a connected
  linear graph $\HH$. Then for every pair of vertices $x,y \in V(\H)$ which
  are incomparable under $\porder_{\H}$, every path $P$ from $x$ to $y$
  satisfies $\min_\H P \not \in \{x,y\}$.
\end{corollary}

\subsubsection*{Reachability and left neighbours, bounded expansion}

\marginpar{$L(\cdot)$}%
Given a tog $\G$, we define the \emph{left neighbourhood} of a vertex $u \in
\G$ as $L(u) := \{v \in N(u) \mid v \porder_\G u\}$. For any integer $r$, we
define the set $\mathcal P^r(u)$ as the set of all paths of length $\leq r$
which have $u$ as one of their endpoints and the set $\mathcal P^r(u,v)$ as
the set of all $u$-$v$-paths of length $\leq r$.  With this notation, we can
now define the \emph{weak $r$-neighbours} as the set
\marginpar{$W^r(\cdot)$}%
\[
  W^r_\GG(u) = \{ \min P \mid P \in \mathcal P^r(u) \},
\]
that is, $W^r_\GG(u)$ contains all vertices that are \emph{weakly $r$-reachable}
from $u$. We also define the \emph{strong $r$-neighbours} as the set
\marginpar{$S^r(\cdot)$}%
\[
  S^r_\GG(u) = \{ v \porder_\G u \mid \exists P \in \mathcal P^r(u,v)
  ~\text{s.t.}~ u \porder_\G (P-v) \},
\]
that is, $S^r_\GG(u)$ contains all vertices that are \emph{strongly $r$-reachable}
from $u$. Note that $W^1$ and $S^1$ are equal to $L$.
\marginpar{$W^r[\cdot{]}$,$S^r[\cdot{]}$}%
For convenience, we define
$W^r_\GG[u] := W^r_\GG(u) \cup \{u\}$ and $S^r_\GG[u] := S^r_\GG(u) \cup \{u\}$.
As usual, we omit the subscript~$\GG$ if clear from the context.

The notions of weak and strong reachability are at the core of the
generalized colourings numbers $\colnum_r$ and $\wcolnum_r$:
\marginpar{$\colnum_r$,$\wcolnum_r$}%
\begin{align*}
  \colnum_r(G) &= \min_{\GG \in \Pi(G)} \max_{v \in G} | W^r_\GG[v] |, \\
  \wcolnum_r(G) &= \min_{\GG \in \Pi(G)} \max_{v \in G} | S^r_\GG[v] |.
\end{align*}
Kierstead and Yang~\cite{WCol} showed that the weak $r$-colouring number is
bounded iff the $r$-colouring number is:
\[
  \colnum_r(G) \leq \wcolnum_r(G) \leq \colnum_r(G)^r
\]
and Zhu related the above graph measures to classes of
bounded expansion~\cite{WColBndExp}. As a result, we can work with the
following characterisation of bounded expansion and nowhere dense
classes:

\begin{proposition}\marginpar{Bounded expansion}
  The following statements about a graph class $\mathcal G$ are equivalent:
  \begin{enumerate}
    \item $\mathcal G$ has \emph{bounded expansion},
    \item there exists a function $f$ such that $\colnum_r(G) < f(r)$
          for all $G \in \mathcal G$ and all $r \in \mathbb N_0$,
    \item there exists a function $g$ such that $\wcolnum_r(G) < g(r)$
          for all $G \in \mathcal G$ and all $r \in \mathbb N_0$.
  \end{enumerate}
\end{proposition}

\noindent
In nowhere dense classes these measures might depend on the size of the
graph, albeit only sublinearly:

\begin{proposition}\marginpar{Nowhere dense}
   The following statements about a graph class $\mathcal G$ are equivalent:
  \begin{enumerate}
    \item $\mathcal G$ is \emph{nowhere dense},
    \item there exists a sequence of functions $(f_r)_{r \in \N_0}$
          with $f_r(n) = O(n^{o(1)})$
          such that $\colnum_r(G) < f_r(|G|)$ for all $G \in \mathcal G$
          and all $r \in \N_0$,
    \item there exists a sequence of functions $(g_r)_{r \in \N_0}$
          with $g_r(n) = O(n^{o(1)})$
          such that $\wcolnum_r(G) < g_r(|G|)$ for all $G \in \mathcal G$
          and all $r \in \N_0$.
  \end{enumerate}
\end{proposition}

\noindent
We are left with the question of computing orderings which provide small
values for $W^r$ or $S^r$. Finding optimal orderings for weakly reachable
sets is \NP-complete~\cite{ColouringCoveringNowhereDense} for~$r \geq 3$,
we therefore have to resort to approximations. The, to our knowledge, best
current option is via \emph{admissibility}, yet another order-based measure:
the $r$-admissibility~$\adm_r^\GG(v)$ of a vertex~$v$ in an ordered graph~$\GG$ is the
maximum number of paths of length at most~$r$ which a) only intersect in~$v$
and b) end in vertices that come before~$v$ in $\leq_\GG$. The admissibility of
a graph~$G$ is then
\begin{align*}
  \adm_r(G) &= \min_{\GG \in \Pi(G)} \max_{v \in G} | \adm_r^\GG (v) |, 
\end{align*}
and it is not too difficult to see that~$\adm_r(\GG) \leq \colnum_r(\GG)$.
In the other direction, we have the following result:
\begin{proposition}[\cf~\Dvorak~\cite{DvorakDomset}]\label{prop:col-adm}
  For any linear ordering~$\GG$ of~$G$ and~$r \in \N$ it holds that
  \[
    \colnum_r(\GG) \leq \adm_r(\GG)(\adm_r(\GG)-1)^{r-1} + 1.
  \]
\end{proposition}

\noindent
Importantly, a linear-time algorithm to compute the admissiblity exists\footnote{
  The algorithm relies on heavy machinery and is in its current formulation
  probably not practical.
}

\begin{proposition}[\cf~\Dvorak~\cite{DvorakDomset}]\label{prop:adm-linear-time}
  Let~$\mathcal G$ be a class with bounded expansion and~$r \in \N$. There
  exists a linear-time algorithm that for each $G \in \mathcal G$ computes
  an ordering~$\GG$ with~$\adm_r(\GG) = \adm_r(G)$.
\end{proposition}

\noindent
As a corollary to these two proposition, we can compute an ordering~$\GG$ of~$G$
in linear time with
\[
  \colnum_r(\GG) \leq \colnum_r(G)(\colnum_r(G)-1)^{r-1} + 1 = O(\colnum_r(G)^r)
\]
and, by applying the result by Kierstead and Yang, with
\[
  \wcolnum_r(\GG) \leq \big(\!\wcolnum_r(G)(\wcolnum_r(G)-1)^{r-1} + 1\big)^r = O((\wcolnum_r(G)+1)^{r^2}).
\]

\subsubsection*{Conventions}

In the remainder, we fix a linear graph $\GG$, the \emph{host graph},
and a \emph{pattern graph} $H$. Our goal is to count how
often $H$ appears as an induced subgraph in the underlying graph $G$
of $\GG$. For ease of presentation, we will assume that $H$ is connected
and discuss later how the algorithms can be modified for disconnected patterns.

\section{Pattern decomposition}\label{sec:patterns}

\noindent
We will be counting the pattern by considering the possible orderings in which
it may appear in the host graph. However, it turns out that some of these
orderings need to be treated as a unit with our approach, namely those
orderings that result in the same pattern relaxation. In that sense, we count
the number of embeddings only for members of the following set:

\begin{definition}[Pattern relaxation]
	For the pattern graph~$H$ we define its \emph{pattern relaxations} as the
	set
	\[
		\mathcal H := \{ \rel((H, \order_\pi)) \mid \pi \in \pi(V(H)) \}.
	\]
\end{definition}

\noindent
Each pattern relaxation will be decomposed further until we arrive at an object
that is easily countable. To that end, we define the following:

\begin{definition}[Pieces, linear pieces]
	Given a pattern relaxation $\H \in \mathcal H$ and a subset
	of its leaves $S \subseteq \leaves(\H)$, the \emph{piece}
	induced by $S$ is the induced subtog
	\[
		\H\big[ \bigcup_{x \in S} \rpath(x) \big].
	\]
  If $|S| = 1$, the resulting piece is a linear graph and
  we refer to it as a \emph{linear piece}.
\end{definition}

\begin{figure}[t]
  \centering
  \includegraphics[scale=.9]{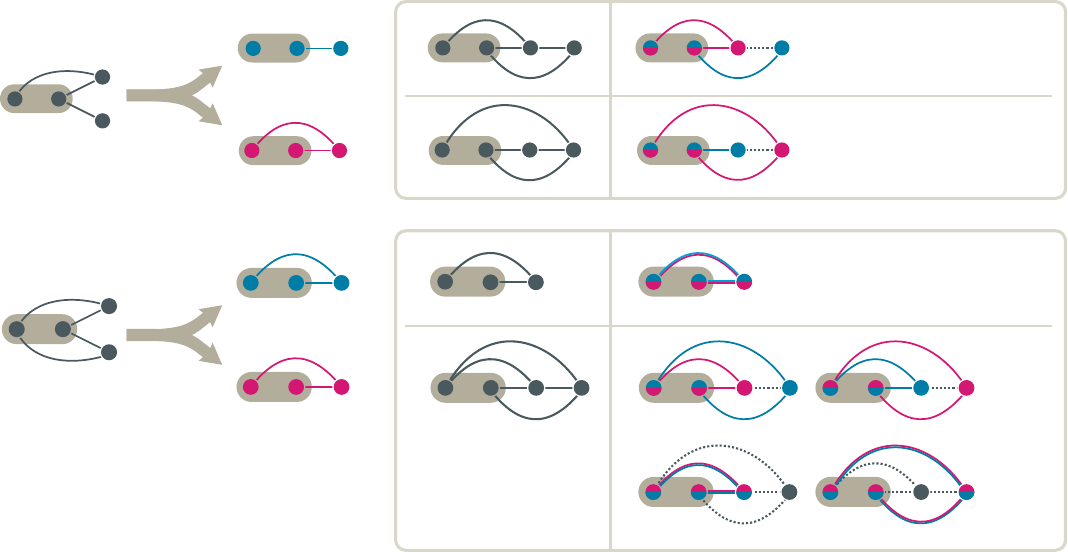}
  \caption{\label{fig:decomp}%
    Piece-sum decompositions for relaxations of a path and a cycle on four
    vertices. The gray highlight around vertices indicates the stem along which
    we decompose. The boxes to the right
    contain all defects (Definition~\ref{def:defect}) for each decomposition (left) as well
    as all embeddings of the two pieces into each defect (right).
  }
\end{figure}

\noindent
With that, we define the decomposition of a pattern relaxation via piece sums
(see Figure~\ref{fig:decomp} for examples):

\begin{definition}[Piece sum]
	Let $\H$ be a tog with stem $\bar x$. We write $\H = \H_1 \oplus_{\bar x} \H_2$
	to denote that $\H_1$ and $\H_2$ are pieces of $\H$ with the property that
	$\leaves(\H_1)$ and $\leaves(\H_2)$ are both non-empty and partition $\leaves(\H)$.
\end{definition}

\noindent
We now show that linear pieces can be enumerated or counted
in linear time given a suitable vertex ordering of the host graph with
constant-sized weak/strong $r$-neighbourhoods.

\subsection*{Counting (relevant) linear pieces}

We first prove that all relevant linear pieces (those that can be completed to
the full pattern) are completely contained in weakly reachable sets
and therefore can be counted easily in time $O(\wcolnum_{|\H|}(\GG)^{\depth (\H)-1} \cdot |\GG|)$, see Section~\ref{sec:algorithms} for details.

\begin{lemma}\label{lemma:linear-count-wcol}
  Let $\KK$ be a linear piece for some pattern relaxation $\H \in \mathcal H$
  and let $z = \max(\KK)$.
  Then for every $\H \embeds{\phi} \GG$
   it holds that $\phi(\KK)$ is contained in $W^{|\H|}_\GG[\phi(z)]$.
\end{lemma}
\begin{proof}
  Let $\phi$ be such an embedding and fix any $x \in \KK$. We need to show
  that $\phi(x) \in W^{|\H|}_\GG[\phi(z)]$. Since we assumed that $H$ is connected, so is
  $\H$. Then by Lemma~\ref{lemma:subtree-path}, there exists a path $P$ path
  from $x$ to $z$ in $\H$ with $\min_\H P = x$. Since $\phi$ is an
  embedding, by Observation~\ref{obs:min-embed} it holds that
  \[
  	\min_\GG \phi(P) = \phi(\min_\H P) = \phi(x).
  \]
  We conclude that $\phi(x) \in W^{|P|}_\GG[\phi(z)] \subseteq W^{|\H|}_\GG[\phi(z)]$.
\end{proof}

\noindent
The above does not hold if we replace weak reachability
by strong reachability, however, the following statement already
suffices to build the strong-reachability variant of our algorithm:

\begin{lemma}\label{lemma:linear-count-col}
  Let $\KK$ be a linear piece for some pattern relaxation $\H \in \mathcal H$.
  Let $z = \max(\KK)$ and let $x <_\KK z$ be an arbitrary vertex of $\KK.$
  There exists a vertex $y \in \KK$, $x <_\KK y \leq_\KK z$ such that for every
  embedding $\H \embeds{\phi} \GG$ it holds that
  $
    \phi(x) \in S_\GG^{|H|}[\phi(y)]
  $.
\end{lemma}
\begin{proof}
  Let $\phi$ be such an embedding. Again, since $\H$ is connected there exists a
  path $P$ from $x$ to $z$ in $\H$ with $\min_\H P = x$. Let $y = \min_\H ((P - x) \cap
  V(\KK))$ be the smallest vertex of $\KK$ which lies on $P$; since $z$ lies in
  this intersection this minimum must exist. Let $P'$ be the portion of $P$
  which goes from $x$ to $y$.

  \begin{claim}
    $y = \min_\H(P'-x)$.
  \end{claim}
  \begin{proof}
    Assume towards a contradiction that $y' = \min_\H(P'-x)$ with $y' \neq y$.
    Note that by our choice of $P$ it holds that $z \sporder_\H y'$.

    First consider the case that $y' \sporder_\H y$.
    Hence $y'$ must lie somewhere on the path from $x$ to $y$ in $T(\H)$.
    But then $y'$ is contained in the piece $\KK$ and hence $(P-x) \cap V(\KK)$,
    contradicting our choice of $y$.

    Otherwise, $y'$ and $y$ are incomparable under
    $\porder_\H$ and in particular $y'$ cannot lie anywhere on $\rpath_{T(\H)}(y)$
    or anywhere below $y$ in $T(\H)$. Since $\porder_\H$ guards $E(H)$ the path
    $P$ can only go from $y'$ to $y$ by intersecting  $\rpath_{T(\H)}(y)$ in
    some vertex $y''$. But then $y'' \in (P-x) \cap V(\KK)$, contradicting our choice
    of $y$.
  \end{proof}
  \noindent
  Finally, we apply Observation~\ref{obs:min-embed} and find that
  \[
    \min_\GG \phi(P'-x) = \phi(\min_\H (P'-x)) = \phi(y)
  \]
  from which we conclude that indeed $\phi(x) \in S^{|P|}_\GG[\phi(y)]
  \subseteq S^{|H|}_\GG[\phi(y)]$.
\end{proof}

\noindent
We call such a vertex $y$ a \emph{hint} and introduce the following
notation to speak about it more succinctly:

\begin{definition}[$\hint$]
  Let $\KK$ be a linear piece of $\H \in \mathcal H$ with
  vertices $x_1,\ldots,x_p$. For every index
  $i \in [p]$ we define the function $\hint^\H_\KK(i)$ to be the largest
  index~$j < i$ such that for every embedding $\H \embeds{\phi} \GG$ it holds that $\phi(x) \in S_\GG^{|H|}[\phi(x_j)]$.
\end{definition}


\subsection*{Combining counts}

In order to succinctly describe our approach, we need to introduce
the following notation for counting embeddings of a pattern graph
$\H$ into a host graph $\G$ where we already fix the embedding
of a prefix of $\H$'s stem vertices.

\begin{definition}[Embedding count]
  For togs $\H, \G$ with $\bar x$ a stem prefix of $\H$ and $\bar y \subseteq \G$
  an ordered vertex with $\bar x = \bar y$, we define
  \[
    \hash_{\xtoy}(\H, \G) := \big|\{ \phi \mid \H \embeds{\phi} \G ~\text{and}~ \phi(\bar x) = \bar y \} \big|.
  \]
\end{definition}

\noindent
The central idea is now that in order to count $\H$, we instead count the
occurrences of two pieces $\H_1 \oplus_{\stem \H} \H_2$ and compute
$\hash_{\stem \H \mapsto \bar y}(\H, \GG)$ by taking the product $\hash_{\stem H
\mapsto \bar  y}(\H_1) \cdot \hash_{\stem H \to \bar  y}(\H_2)$. Of course, the
latter quantity overcounts the former, as we will discuss below. First,
let us introduce the following notation for this `estimate' embeddings
count:

\begin{definition}[Relaxed embedding count]
  For togs $\H = \H_1 \oplus_{\bar x} \H_2, \G$
  with $\bar x$ a stem prefix of $\H$ and $\bar y \subseteq \G$
  an ordered vertex set with $\bar x = \bar y$, we define
  \[
    \hash_{\xtoy}(\H, \G \mid \H_1, \H_2) :=
     \big|\{ \phi \in V(\G)^{V(\H)} \mid \H_1 \embeds{\phi} \G, \H_2 \embeds{\phi} \G ~\text{and}~ \phi(\bar x) = \bar y \} \big|.
  \]
\end{definition}

\noindent Now, how does a pair of embeddings for $\H_1$, $\H_2$ fail to embed
$\H$? We either must have that the embeddings' images intersect or that there
exists an edge between their images which is `invisible' to the individual
embeddings. We will call such pair of embeddings  a \emph{defect}:

\begin{definition}[Defect]\label{def:defect}
	Let $\H \in \mathcal H$ be a pattern relaxation and let $\H_1 \oplus_{\bar
	x} \H_2 = \H$ with $\bar x = \stem \H$. A \emph{defect} of $\H_1, \H_2$ is any etog $\D$
	that satisfies the following properties:
	\begin{enumerate}
		\item $\H \notembeds \D$,
		\item $\H_1 \embeds{id} \D$,
		\item $\H_2 \embeds{\phi} \D$ where $\phi$ is the identity
			 on the set $V(\H_2)\setminus V(\H_1) \cup \bar x$,
		\item and $V(\D) = V(\H_1) \cup \phi(\H_2)$.
	\end{enumerate}
	We will write $\defects(\H_1,\H_2)$ to denote the set of all defects for the
	pair $\H_1,\H_2$.
\end{definition}

\noindent
Note that several of the above properties are for convenience only: we
insist that $\H_1$ is a subgraph of $\D$ to avoid handling yet another embedding
and we make $\phi$ preserve all vertices that it possibly can for the same
reason. Importantly, all the togs $\H$, $\H_1$, $\H_2$, and $\D$ share the ordered
set $\bar x$ as a stem prefix.

At this point we should point out that it is not a priori clear that it is enough
to consider defects that are etogs themselves, it could very well be the case that
defects are arbitrary tree-ordered or just `ordered' graphs. Note that what we
\emph{really} want to count are linear subgraphs $\DD \subgraph \GG$ into
which $\H_1$ and $\H_2$ embed, but $\H = \H_1 \oplus_{\bar x} \H_2$ does not
(as these are precisely the cases that we overcount in the product
$\hash_{\bar x \mapsto \bar  y}(\H_1) \cdot \hash_{\bar x \mapsto \bar  y}(\H_2)$,
for some prefix $\bar y$ of $\DD$). The next lemma shows that instead of
trying to find these linear subgraphs, we can instead recourse to counting
their relaxations, thus circling back to etogs:

\begin{lemma}\label{lemma:frankie-says-relax}
  Let $\H$ be a connected etog with pieces $\H_1 \oplus_{\bar x} \H_2 = \H$. Let
  $\D \in \defects(\H_1,\H_2)$. Then for every linear graph $\DD$ with $\rel(\DD) = \D$
  it holds that
  \[
    \hash_{\xtox}(\H, \D \mid \H_1, \H_2) = \hash_{\xtox}(\D, \DD) \hash_{\xtox}(\H, \DD \mid \H_1, \H_2)
  \]
\end{lemma}
\begin{proof}
  First consider any mapping $\phi \in V(\D)^{V(\H)}$ with $\H_i \embeds{\phi} \D$ for $i \in \{1,2\}$
  and $\phi(\bar x) = \bar x$. Then for every embedding $\D \embeds{\psi} \DD$
  with fixed points $\bar x$, it holds that $\H_i \embeds{\psi \circ \phi} \DD$,
  hence
  \[
    \hash_{\xtox}(\H, \D \mid \H_1, \H_2)
    \leq \hash_{\xtox}(\D, \DD) \hash_{\xtox}(\H, \DD \mid \H_1, \H_2).
  \]
  For the other direction, consider any mapping $\theta \in V(\DD)^{V(\H)}$
  with $\H_i \embeds{\theta} \DD$ for $i \in \{1,2\}$. Let again $\D
  \embeds{\psi} \DD$ with fixed points $\psi(\bar x)$. Since $\D =
  \relax(\DD)$, $\psi$ is a bijection and the mapping $\xi := \psi^{-1} \circ
  \theta$ from $V(\H)$ to $V(\D)$ is well-defined.

  \begin{claim}
    $\H_i \embeds{\xi} \D$ for $i \in \{1,2\}$.
  \end{claim}
  \begin{proof}
    It is easy to see that $\xi$
    preserves the edge relations of $\H_1$ and $\H_2$, therefore let us
    consider the ordering relations.

    Since $\bar x$ are fixed points for all
    the mappings involved we can conclude that the ordering relation with
    respect to pairs with at least one member in $\bar x$ is preserved by $\xi$.
    Therefore, consider $u, v \in \H_1 - \bar x$ (the argument for pairs in
    $\H_2 - \bar x$ is the same) with $u \porder_{\H_1} v$. By Lemma~\ref{lemma:subtree-path},
    there exists a $u$-$v$ path $P$ in $\H$ with $\min_\H P = u$. It follows
    that $P \subseteq V(\H_1)$ and therefore $\min_{\H_1} P = u$. Since
    $\H_1 \embeds{\theta} \DD$, by Observation~\ref{obs:min-embed}
    we have that $\min_{\DD} \theta(P) = \theta(u)$.

    Assume towards a contradiction that $\xi(u) \not \porder_{\D} \xi(v)$. If
    $\xi(v) \porder_{\D} \xi(u)$ then $\theta(v) \porder_{\DD} \theta(u)$,
    contradicting $\H_1 \embeds{\theta} \DD$. Therefore assume that $\xi(u),
    \xi(v)$ are incomparable under $\porder_{\D}$. Note that, since $P \subseteq V(\H_1)$,
    $\xi(P)$ is a $\xi(u)$-$\xi(v)$-path in $\D$. Because the two endpoints
    are assumed to be incomparable, by Corollary~\ref{cor:incomparable-path}
    it follows that $\min_\D \xi(P) \not \in \{ \xi(u), \xi(v) \}$. By
    Observation~\ref{obs:min-embed}, it follows that
    \begin{align*}
      \psi\big(\min_\D \xi(P)\big) \not \in \big\{ \psi(\xi(u)), \psi(\xi(v)) \big\}
      &\implies
      \min_\DD \psi(\xi(P)) \not \in \big\{ \psi(\xi(u)), \psi(\xi(v)) \big\} \\
      &\implies
      \min_\DD \theta(P) \not \in \{ \theta(u), \theta(v) \}
    \end{align*}
    This of course contradicts our earlier conclusion that $\min_{\DD}
    \theta(P) = \theta(u)$ and we conclude that the claim holds.
  \end{proof}

  \noindent
  Our construction of $\xi$ works for any $\D \embeds{\psi} \DD$ with
  fixed points $\bar x$, of which there are $\hash_{\xtox}(\D, \DD)$
  many. Therefore
  \[
    \hash_{\xtox}(\D, \DD) \hash_{\xtox}(\H, \DD \mid \H_1, \H_2)
    \leq
    \hash_{\xtox}(\H, \D \mid \H_1, \H_2).
  \]
  Taking both inequalities together, we conclude that the lemma holds.
\end{proof}

\noindent
Lemma~\ref{lemma:frankie-says-relax} still leaves us with the awkward
quantity $\hash_\xtox(\D,\DD)$, again, we would like to compute with etogs
and not linear graphs. The following lemma shows that indeed we only
really need to count rooted automorphisms of etogs:

\begin{lemma}\label{lemma:auto-linear}
  Let $\H$ be a connected etog with stem prefix $\bar x$
  and let $\HH$ be a linear graph with $\rel(\HH) = \H$.
  Then $\hash_\xtox(\H, \H) = \hash_\xtox(\H, \HH)$.
\end{lemma}
\begin{proof}
  Fix a single embedding $\H \embeds{\phi} \HH$ with $\phi(\bar x) = \bar x$,
  as observed above such an embedding always exists. Note that $\phi$ is
  necessarily a bijection.

  In the first direction, let $\xi$ be an $\H$-automorphism with fixed points
  $\bar x$, then $\H \embeds{\phi \circ \xi} \HH$ as well. Since this
  construction yields a unique mapping for every such automorphism, we
  conclude that $\hash_\xtox(\H,\H) \leq \hash_\xtox(\H,\HH)$.

  In the other direction, consider any embedding $\H \embeds{\psi} \HH$
  with $\psi(\bar x) = \bar x$,
  again, $\psi$ is a bijection. Define the mapping $\theta = \psi^{-1} \circ \phi$,
  we claim that $\theta$ is an $\H$-automorphism with fixed points $\bar x$.
  Since $\phi(\bar x) = \bar x = \psi(\bar x)$, the latter part follows immediately.
  That $\theta$ preserves the edge relationship of $\H$ also follows easily,
  we are left to argue that the order relationship is preserved.

  Let $u \porder_{\H} v$ and assume towards a contradiction that $\theta(u)
  \not \porder_{\H} \theta(v)$. If $\theta(v) \porder_{\H} \theta(u)$ we find
  a contradiction because $\phi(u) \porder_{\HH} \phi(v)$ but also $\H
  \embeds{\psi} \HH$, which implies $\psi(\theta(v)) \porder_{\HH}
  \psi(\theta(u))$ and thus $\phi(v) \porder_{\HH} \phi(u)$. Hence we are
  left with the case that $\phi(u)$, $\phi(v)$ are incomparable under
  $\porder_{\H}$. Since $\H$ is connected, by Lemma~\ref{lemma:subtree-path}
  there exists a $u$-$v$-path $P$ in $\H$ with $\min_\H P = u$, and by
  Observation~\ref{obs:min-embed} we have that $\min_\HH \phi(P) = \phi(u)$.
  Now consider the $\theta(u)$-$\theta(v)$-path $\theta(P)$ in $\H$: since we
  assumed $\theta(u)$ and $\theta(v)$ to be incomparable under $\porder_\H$,
  by Corollary~\ref{cor:incomparable-path} it holds that $\min_\H \theta(P)
  \neq \theta(u)$. But by Observation~\ref{obs:min-embed},
  \[
    \psi(\min_\H \theta(P)) \neq \psi(\theta(u))
    \implies
    \min_\HH \phi(P) \neq \phi(u).
  \]
  We arrive at a contradiction and conclude that $\theta$ is indeed an
  $\H$-automorphism and therefore $\hash_\xtox(\H,\HH) \leq \hash_\xtox(\H,\H)$.

  Hence, it actually holds that $\hash_\xtox(\H,\H) = \hash_\xtox(\H,\HH)$,
  as claimed.
\end{proof}

\noindent
We are now ready to prove the main technical lemma of this paper, the recurrence that will
allow us to compute $\hash_{\xtoy}(\H,\GG)$, \ie the number of embeddings from
$\H$ into $\GG$ which map the stem prefix $\bar x$ of $\H$ onto the ordered
subset $\bar y$ of $\GG$. Note that in order to compute the number of
\emph{induced subgraphs}, we simply have to divide this value by
$\hash_{\xtox}(\H, \H)$, the number of automorphisms of $\H$ with fixed points $\bar x$.

\begin{lemma}\label{lemma:counting}
	Let $\H \in \mathcal H$ be a (non-linear) pattern relaxation and let $\H_1 \oplus_{\bar
	x} \H_2 = \H$. Fix an ordered vertex set $\bar y \in \GG$ such that
	$\H[\bar x] \isom \GG[\bar y]$. Then
	\begin{align*}
		 \hash_{\xtoy}(\H,\GG) =
     \! \hash_{\xtoy}(\H_1, \GG) \! \hash_{\xtoy}(\H_2, \GG)
			- \mkern-16mu \sum_{\!\D \in \defects(\H_1,\H_2)}
					\frac{\displaystyle   \hash_{\xtox}(\H, \D \mid \H_1,\H_2) \hash_{\xtoy}(\D, \GG)}
               {\displaystyle   \hash_{\xtox}(\D, \D)^2}.
	\end{align*}
\end{lemma}
\begin{proof}
	Let $\HH \subseteq \GG$ be a linear subtog whose stem has the prefix $\bar y$.
	Let $\Phi(\HH)$ contain all pairs $(\phi_1,\phi_2)$ with $\phi_1(\bar x ) = \phi_2(\bar x) = \bar y$
  such that
  \begin{enumerate}[a)]
    \item $\H_1 \embeds{\phi_1} \HH$ and $\H_2 \embeds{\phi_2} \HH$; and
    \item $V(\HH) = V(\phi_1(\H_1)) \cup V(\phi_2(\H_2))$.
  \end{enumerate}
	That is, $\Phi(\HH)$
	contains all pairs of embeddings that minimally embed the graphs $\H_1$
	and $\H_2$ into $\HH$. Note that every pair of embeddings appears in
	precisely one such set, namely $\Phi(\GG[V(\phi_1(\H_1)) \cup V(\phi_1(\H_2))])$.
  Accordingly,
  \[
    \hash_{\xtoy}(\H_1, \GG) \! \hash_{\xtoy}(\H_2, \GG)
    = \sum_{\HH \subtog \GG} |\Phi(\HH)|.
  \]
	We will therefore count how embedding-pairs contribute to the above product
  by arguing about their $\Phi$-associated subtog~$\HH$.
  We distinguish the following cases:

	\medskip
	\noindent\textbf{Case 1:} $\H \embeds{\xtoy\,} \HH$. \\
	It follows that $\rel(\HH) \not \in \defects(H_1,H_2)$, therefore
	$\Phi(\HH)$ does not contribute to the sum on the right-hand side.
	Every pair in $\Phi(\HH)$ is counted by the product on the right-hand side,
	therefore we have to argue that they are counted in the left-hand side as well.

	$\HH$ contributes to $\hash_{\bar x \to \bar y}(\H, \GG)$ an amount of
	$\hash_{\bar x \to \bar y}(\H, \HH)$, so the following equation must hold
	in order for $\Phi(\HH)$ to contribute equal amounts on both sides:
	\[
	 \hash_{\bar x \to \bar y}(\H, \HH)	= |\Phi(\HH)|.
	\]
	Fix any pair $(\phi_1,\phi_2) \in \Phi(\HH)$.
	Since $\H \embeds{~~} \HH$ and $V(\HH) = V(\phi_1(\H_1)) \cup V(\phi_1(\H_2))$, we
	conclude that $|\HH| = |\H|$. Furthermore, the sets $V_1 := V(\phi_1(\H_1))$ and $V_2 := V(\phi_2(\H_2))$
	must be disjoint and therefore partition $V(\HH)$.

	\begin{claim}
    Define the mapping $\phi\colon V(\H) \to V(\HH)$ as
  	$\phi(u) = \phi_1(u)$ for all $u \in V(\H_1)$ and $\phi(u) = \phi_2(u)$
  	for all other vertices.	Then $\H \embeds{~\phi~} \HH$.
	\end{claim}
	\begin{proof}
		Clearly, $\phi(\bar x) = \bar y$, thus we can focus
		on the non-stem parts of $\H$. Now, the only reason why $\phi$ would \emph{not} be an embedding
		is if there exists an edge between $V_1$ and $V_2$ in $\HH$: we already
		established that $V_1 \cap V_2 = \emptyset$, and since $\H$ is a relaxation and
		there are no edges between $\H_1 - \bar x$ and $\H_2 - \bar x$ it follows that the
		relative order of these two sides is unconstrained by $\porder_{\H}$. We easily
		arrive at a contradiction by counting edges: the pieces $\H_1$ and $\H_2$ cover
		all edges of $\H$, hence an additional edge between $V_1$ and $V_2$ would contradict
		that $\H \embeds{~~} \HH$.
  \end{proof}

  \noindent
  The above claim shows that there is a one-to-one correspondence between
  embeddings of $\H$ into $\HH$ and the embedding-pairs in $\Phi(\HH)$,
  and therefore $\hash_{\bar x \to \bar y}(\H,\HH) = |\Phi(\HH)|$.

	\smallskip
	\noindent\textbf{Case 2:} $\H \centernot{\embeds{\xtoy\,}} \HH$. \\
  First assume that $\Phi(\HH)$ is non-empty, therefore we can conclude that
  $\rel(\HH) \in \defects(\H_1,\H_2)$. Because
  $\H \centernot{\embeds{\xtoy\,}} \HH$, the pairs $\Phi(\HH)$
  do not contribute to the left-hand side and we are left with showing that each
  such pair is subtracted by the sum on the right-hand side. The relevant term
  here is, of course, when the sum index takes on the value $\D := \rel(\HH)$:
  \[
    \frac{\hash_{\xtox}(\H, \D \mid \H_1,\H_2) \hash_{\xtoy}(\D, \GG)}
         {\hash_{\xtox}(\D, \D)^2}
  \]
  Note that the term $\hash_{\xtoy}(\D, \GG)/\hash_{\xtox}(\D, \D)$
  counts the number of subtogs $\DD \subtog \GG$ with stem prefix $\bar y$ into
  which $\D$ embeds, \ie $\D \embeds{\xtoy} \DD$. Accordingly, $\HH$
  contributes precisely one to this term and we are left to show that
  \begin{align}
    \frac{\hash_{\xtox}(\H, \D \mid \H_1,\H_2)}{\hash_{\xtox}(\D, \D)} = |\Phi(\HH)|. \label{eq:frac}
  \end{align}
  To that end, we first prove the following:

  \begin{claim}
   $|\Phi(\HH)| = \hash_{\xtoy}(\H,\HH \mid \H_1, \H_2)$.
  \end{claim}
  \begin{proof}
    In the one direction, fix a pair $(\phi_1, \phi_2) \in \Phi(\HH)$. Define
    the mapping $\phi$ via $\phi(u) = \phi_1(u)$ for $u \in \H_1$ and $\phi(u) = \phi_2(u)$
    otherwise. Since $V(\H_1) \cap V(\H_2) = \bar x$ and $\phi(\bar x) = \bar y$,
    we conclude from $\H_1 \embeds{\phi_1} \HH$ and $\H_1 \embeds{\phi_2} \HH$
    that indeed $\H_1 \embeds{\phi} \HH$ and $\H_2 \embeds{\phi} \HH$.
    Since $V(\H) = V(\H_1) \cup V(\H_2)$ and $V(\HH) = V(\phi_1(\H_1)) \cup V(\phi_2(\H_2))$,
    we also have that $\phi \in V(\HH)^{V(\H)}$,
    thus $|\Phi(\HH)| \leq \hash_{\xtoy}(\H,\HH \mid \H_1, \H_2)$.

    In the other direction, take a mapping $\phi \in V(\HH)^{V(\H)}$
    with $\H_1 \embeds{\phi} \HH$, $\H_2 \embeds{\phi} \HH$ and $\phi(\bar x) = \bar y$.
    Define $\phi_i := \phi|_{V(\H_i)}$, then $\H_i \embeds{\phi_i} \HH$
    for $i \in \{1,2\}$,
    thus $\hash_{\xtoy}(\H,\HH \mid \H_1, \H_2) \leq |\Phi(\HH)|$.
    We conclude that the two sides are actually equal, as claimed.
  \end{proof}
  \noindent Equation~\ref{eq:frac} is thus equivalent to
  \[
    \hash_{\xtox}(\H, \D \mid \H_1,\H_2) = \hash_{\xtox}(\D, \D) \hash_{\xtoy}(\H,\HH \mid \H_1, \H_2).
  \]
  Applying Lemma~\ref{lemma:auto-linear} gives this is equivalent to
  \[
    \hash_{\xtox}(\H, \D \mid \H_1,\H_2) = \hash_{\xtox}(\D, \DD) \hash_{\xtoy}(\H,\HH \mid \H_1, \H_2),
  \]
  which of course is precisely the statement of Lemma~\ref{lemma:frankie-says-relax}.

  We conclude that the two sides of the equation are indeed equal, as claimed.
\end{proof}

\noindent

\begin{GrayBox}{\textbf{Note}}
  For practical purposes, it is preferable to compute embedding-counts which exclude automorphisms.  Define~$\thash_{\xtoy}(\H,\GG) := \hash_{\xtoy}(\H,\GG) / \hash_{\xtox}(\H,\H)$ to be this automorphism-corrected
  count, then the equation in Lemma~\ref{lemma:counting} becomes
  \begin{align*}
     \thash_{\xtoy}(\H,\GG) &=
      \frac{\hash_{\xtox}(\H_1, \H_1) \hash_{\xtox}(\H_2, \H_2)}{\hash_{\xtox}(\H, \H)} \thash_{\xtoy}(\H_1, \GG) \! \thash_{\xtoy}(\H_2, \GG)  \\
      & - \mkern-16mu \sum_{\!\D \in \defects(\H_1,\H_2)}
          \frac{\displaystyle   \hash_{\xtox}(\H, \D \mid \H_1,\H_2)}
               {\displaystyle  \hash_{\xtox}(\H, \H)  \hash_{\xtox}(\D, \D)} \thash_{\xtoy}(\D, \GG).
  \end{align*}
  The above form is better suited for implementation as the numbers stay smaller but for the mathematical presentation the form in
  Lemma~\ref{lemma:counting} is simpler.
\end{GrayBox}

\noindent
We next prove that the recurrence implied by the equation in Lemma~\ref{lemma:counting} ends after only a few steps.

\begin{lemma}\label{lemma:recurrence-depth}
  The recurrence for $\hash_{\xtoy}(\H,\GG)$ as stated in
  Lemma~\ref{lemma:counting} has depth at most $|\H|$.
\end{lemma}
\begin{proof}
	We argue that the measure $f(\G) = |\G| - |\stem(\G)|$ strictly decreases
	for all graphs involved on the right hand side, \ie $f(\K) < f(\H)$ for
	all graphs $\K \in \{ H_1, H_2 \} \cup \defects(H_1,H_2)$. Since $H_1
	\oplus_{\bar x} \H_2 = \H$, both togs $H_1$ and $H_2$ are proper pieces of
	if $\H$ and we conclude that $|\H_i| < |\H|$ and $|\stem(H_i)| > |\stem(H)|$,
	thus $f(\H_i) < f(\H)$ for $i \in \{1,2\}$.

	We are left to prove the same for $\D \in \defects(\H_1,\H_2)$. Let~$\phi$ be the embedding
	$\H_2 \embeds{~\phi~} \D$. Since $\D \not \isom \H$ but $V(\D) = V(\H_1) \cup \phi(\H_2)$,
	we conclude that $\H_1 \setminus \bar x$ and $\phi(\H_2) \setminus \bar x$ must either
	be connected by an edge or share a vertex in $\D$. In either case, the subtog
	$\D - \bar x$ is connected, hence $|\stem(\D)| > |\stem(\H)|$ while
	$|\D| \leq |\H|$. We conclude that indeed $f(\D) < f(\H)$.

	Finally, note that if $f(\G) = 0$ then the tog $\G$ is linear, hence the recurrence
	ends after at most $f(\H) \leq |\H|$ steps.
\end{proof}

\subsection*{Computing defects}

For the remainder of this section, fix $\H = \H_1 \oplus_{\bar x} \H_2$ where
$\bar x := \stem(\H)$. Let also $V_1 := V(\H_1) - \bar x$ and $V_2 := V(\H_2)
- \bar x$ be the vertex sets exclusive to $\H_1$ and $\H_2$.

\begin{definition}[Monotone]
  Let $\porder$ be a partial order over a set $S$ and let $M \subseteq {S \choose 2}$  be a matching.
  Let further $\dir D$ be the digraph representation of $\porder$.
  We say that $M$ is \emph{monotone} with respect to $\porder$ if the
  digraph obtained from $\dir D$ by identifying the pairs
  in $M$ is a dag.
\end{definition}

\begin{definition}[Defect map]
  A \emph{defect map} is a bijection $\kappa\colon \bar x \cup \tilde V_1
  \to \bar x \cup \tilde V_2$ for subsets $\tilde V_1 \subseteq V_1$
  and $\tilde V_2 \subseteq V_2$  with the following  properties:
  \begin{itemize}
      \item $\kappa$ is an isomorphism between $H[\bar x \cup \tilde V_1]$
            and $H[\bar x \cup \tilde V_2]$,
      \item the matching $\{x\kappa(x) \mid x \in \tilde V_1\}$ is
            monotone with respect to $\porder_\H$.
  \end{itemize}
\end{definition}

\noindent
In the following we construct a set of etogs $\mathcal D'$ and prove that it
is precisely $\defects(\H_1,\H_2)$. Given the decomposition $\H_1 \oplus_{\bar x} \H_2$
of $\H$, we generate the etogs in $\mathcal D'$ as follows:

\begin{enumerate}
    \item Select appropriate subsets $\tilde V_1 \subseteq V_1$
          and $\tilde V_2 \subseteq V_2$ and a defect map $\kappa\colon \bar x \cup \tilde V_1 \to \bar x \cup \tilde V_2$. Let $M := \{x\kappa(x) \mid
          x \in \tilde V_1 \}$.
    \item Identify the pairs matched by $M$ in $H$
          to create the (unordered) graph $H'$ and create
          the relation $\porder_M$ from $\porder_\H$ by the same process.
    \item Select a set $E^+ \subseteq (V_1 - \tilde V_1) \times (V_2 - \tilde V_2)$
          with $E^+ \cap E(H) = \emptyset$ and add it to $H'$; we
          only allow $E^+ = \emptyset$ if $\tilde V_1, \tilde V_2 \neq \emptyset$.
    \item For every linear ordering $\leq$ of $V(H)$ that is compatible with $\porder_M$,
          add the graph $\rel((H',\leq))$ to $\mathcal D'$.
\end{enumerate}

\noindent
For compatibility with Definition~\ref{def:defect}, whenever we
identify vertices $xy \in M$, we label the resultant vertex $x$,
thus $V_1 \subseteq V(H')$.

\begin{theorem}\label{thm:generate-defects}
  The above process generates exactly $\rel(\H_1,\H_2)$.
\end{theorem}

\noindent
We prove Theorem~\ref{thm:generate-defects} by showing the following
two lemmas.

\begin{lemma}
  $\defects(\H_1,\H_2) \subseteq \mathcal D'$.
\end{lemma}
\begin{proof}
  Consider $\D \in \defects(\H_1,\H_2)$ and let $\H_2 \embeds{\phi} \D$ such
  that $\phi$ is the identity on the set $V(\H_2)\setminus V(\H_1) \cup \bar
  x$. Recall that, by convention, $\H_1 \embeds{id} \D$. Let $\tilde V := V_1
  \cap \phi(V_2)$ and define the mapping $\kappa\colon \bar x \cup \tilde V
  \to \bar x \cup \phi^{-1}(\tilde V)$ as the identity on $\bar x$ and
  $\kappa(x) := \phi^{-1}(x)$ for $x \in \tilde V \subseteq V_1$.
  Let further $M := \{x\kappa(x) \mid x \in \tilde V\}$.

  \begin{landscape}
  \begin{figure}
    \centering\includegraphics[scale=.8]{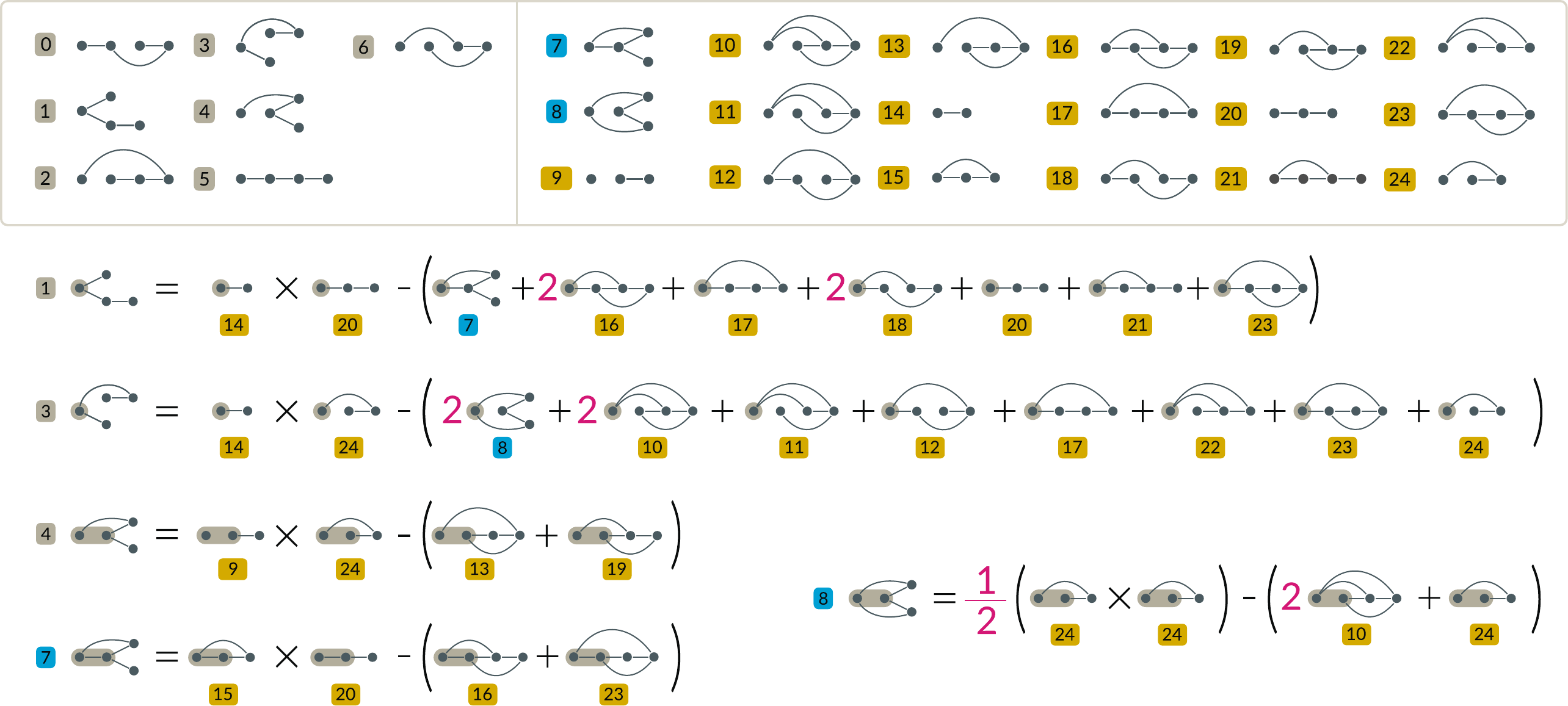}
    \caption{\label{fig:P4}%
      \newline
      \textbf{Top}: Patterns (0--6), defects (7--8) and pieces (9--24) needed to count a path on four vertices. The arrangements indicate the tree order, all pattern relaxations except 1,3,4,7 and 8 are linear.
      \newline
      \textbf{Bottom}: Algebraic expressions to compute non-linear patterns,
      gray boxes indicate the stems.
      The graphs are understood as automorphism-corrected embedding counts
      $\thash_{\xtoy}(\H,\GG)$ (see note below Lemma~\ref{lemma:counting}).  For example, in order
      to compute the number of embeddings of defect~8 which map its stem
      $\bar x$ onto a vertex pair $\bar y$ in the host graph, we first need to compute
      the number of embeddings of the pieces~24 and~10 which likewise map the
      first two vertices of their stem prefix onto $\bar x$.
    }
  \end{figure}
  \end{landscape}

  \begin{claim}
    $\kappa$ is a defect map.
  \end{claim}
  \begin{proof}
    Since $\H_1[\bar x \cup \tilde V] \embeds{id} \H[\bar x \cup \tilde V]$
    and $\H_2[\kappa(\bar x \cup \tilde V)] \embeds{\kappa^{-1}} \H[\bar x
    \cup \tilde V]$ we conclude that $\kappa$ is an isomorphism of the
    underlying graphs $H_1[\bar x \cup \tilde V]$ and $H_2[\kappa(\bar x \cup
    \tilde V)]$.

    Let $\dir O$ be the digraph representation of $\porder_\H$ and $\dir O''$
    of $\porder_\D$. By construction, $\dir O''$ is precisely the digraph
    obtained from $\dir O$ by identifying the pairs matched in $M$. Since $\D$
    is a tree-ordered graph, $\porder_\D$ is a partial order and thus $\dir
    O''$ is a dag. In other words, the matching $M$ is monotone with respect
    to $\porder_\H$ and we conclude that $\kappa$ is a defect map.
  \end{proof}

  \noindent
  Let $\tilde V_1 := \tilde V$ and $\tilde V_2 := \kappa(\tilde V)$ in
  the following.  Define $E^+ := E(\D)
  \cap ((V_1 - \tilde V_1)\times(\phi(V_2) - \tilde V_1))$.
  Let $H'$ be the graph obtained from $H$ by identifying the pairs matched by
  $M$ and adding $E^+$ to it. Let further $\porder_M$ be the relation obtained
  from $\porder_{\H}$ by identifying the pairs matched by $M$. It is left to
  show that there exists a linear order $\leq$ which is compatible with
  $\porder_M$ and satisfies $D = \rel((H',\leq))$. Let $\dir O'$ be the
  digraph representation of $\porder_M$ and let $\dir O''$ be again the
  digraph representation of $\porder_\D$. Note that the difference between
  $\dir O'$ and $\dir O''$ are arcs corresponding to an orientation $\dir E^+$
  of $E^+$ and transitive arcs resulting from the addition of $\dir E^+$.
  Since all edges in $E^+$ are between $V_1 - \tilde V$ and $\phi(V_2) -
  \tilde V$ and those two sets are disjoint, we can choose, for example,
  to orient $\dir E^+$ be letting all arcs point towards $\phi(V_2) - \tilde V$.
  Then $\dir O' \cup \dir E^+$ is a digraph and so is its transitive
  closure $\tilde O''$. Now note that every topological
  ordering~$\leq$ of $\dir O''$ is also a topological ordering of $\dir O'$
  and we conclude that $\porder_M$ is compatible with $\porder_D$. Since
  $\D$ is an etog, these orderings also all satisfy $\rel((H',\leq)) = D$.
  We conclude that $D \in \mathcal D'$ and therefore $\defects(\H_1,\H_2) \subseteq \mathcal D'$.
\end{proof}

\begin{lemma}
  $\mathcal D' \subseteq \defects(\H_1,\H_2)$.
\end{lemma}
\begin{proof}
  Consider $\D \in \mathcal D'$ and let $\tilde V_1$, $\tilde V_2$,
  $\kappa$, $E^+$ and $\leq$ be those choices that generated $\D$.
  Let also $H'$ be the graph generated by identifying the pairs
  matched by $M := \{x\kappa(x) \mid x \in \tilde V_1\}$ in $H$
  and adding $E^+$ to the resulting graph. We need to show that
  $\D$ is indeed a defect; note that by the last step of the
  construction it is necessarily an etog.

  First, let us convince ourselves that $\H \notembeds \D$. If
  $\tilde V_1 \neq \emptyset$, then $\D$ has less vertices than $\H$
  and thus no embedding can exist. Otherwise, we have that $E^+$
  is non-empty and therefore $\D$ has more edges than $\H$, again
  no embedding can exist. \looseness-1

  Next, we need to show that $\H_1 \embeds{id} \D$. We chose to
  label the vertices from identifying the pairs in $M$ by their
  respective endpoint in $\tilde V_1$. Furthermore, no edge in $E^+$
  has both its endpoints in $\bar x \cup V_1$, therefore $\D[\bar x \cup V_1]
  = \H[\bar x \cup V_1]$ and therefore $\H_1 \embeds{id} \D$.

  Similarly, we need to show that $\H_2 \embeds{\phi} \D$. Define
  $\phi$ to be the identity on $\bar x \cup V_2 \setminus V_1$
  and $\kappa^{-1}$ on $\tilde V_2$. Again, no edge in
  $E^+$ has both its endpoints in $\phi(\H_2)$ and hence
  $\H[V_2] \embeds{id} \D[\phi(\H_2)]$ and therefore
   $\H_2 \embeds{\phi} \D$.

  Finally, it follows directly from the construction of $\D$
  that indeed $V(\D) = \bar x \cup V_1 \cup \phi(V_2) = V(\H_1) \cup \phi(\H_2)$,
  thus we conclude that $\D$ is indeed a defect. It follows that
  $\mathcal D' \subseteq \defects(\H_1,\H_2)$, as claimed.
\end{proof}

\section{The algorithms}\label{sec:algorithms}

In order to efficiently implement the counting algorithm we need a data
structure $\CC$ which acts as a map from ordered vertex sets to integers; the
idea being that for a fixed pattern relaxation $\H \in \mathcal H$ with stem
$\bar x$ we store in $\CC[\bar y]$, $\bar y \subset \GG$ how many embeddings $\H
\embeds{\bar x \mapsto \bar y} \GG$ exist. We use
Lemma~\ref{lemma:linear-count-wcol} or Lemma~\ref{lemma:linear-count-col} to
populate these counters for all linear pieces of $\H$ and then use
Lemma~\ref{lemma:counting} to progressively compute counts for larger and larger
pieces of $\H$ until we arrive at a count for $\H$ itself. We organize the
progressive decompositions of $\H$ and the coefficients resulting from the
application of Lemma~\ref{lemma:counting} in a \emph{counting dag}. Leaves of the
counting dag correspond to linear pieces of $\H$, the single source to $\H$ itself.
The computation then proceeds from the leaves upwards; a task can be completed
as soon as all its out-neighbours have been completed  (leaf nodes are
completed by applying Lemma~\ref{lemma:linear-count-wcol}
or Lemma~\ref{lemma:linear-count-col}).

Repeating
this procedure for all pattern relaxations in $\mathcal H$ and correcting the
sum by the number of automorphisms of $H$ then gives us the total number of
times $H$ appears as an induced subgraph of $G$. For convenience, we compute
a joint counting dag for all relaxations $\mathcal H$ and read of the final value
from all its source nodes---note that in practice this will save some computations
since the counting dags likely have nodes in common.

We first outline the notation and necessary operations of $\CC$ and then
discuss how it can be implemented, then we describe the counting dag and then
finally provide the algorithms. We will assume in the following that $\GG$
is a linear ordering of $G$, we present the algorithm with a dependence on
$\wcolnum_{|H|}(G)$ and show what modifications have to be made to arrive
at an algorithm depending on $\colnum_{|H|}(G)$ instead.

\subsection*{Counting data structure}

The \emph{counting data structure $\CC$ of depth $d$} is a map from $d$-length
ordered vertex sets $\bar y \subseteq \G$ to positive integers $\CC[\bar y]$.
Initially, the counting data structure contains a count of zero for every possible key.
We write $|\CC|$ to denote the number of keys stored in $\CC$ with non-zero counts.
The data structure supports the following queries and modifications:
\begin{itemize}
  \item \emph{Increment} count~$\CC[\bar y]$ by any integer for tuples $\bar y$ of
        length $d$ in time $O(d)$;
  \item Answer the \emph{prefix query}
        \[
            \CC[\bar y] := \sum_{\substack{\bar z: |\bar z| = d \\ \text{and}~ \bar z|_ r= \bar y}} \CC[\bar z]
        \]
        for tuples $\bar y$ of length $r \leq d$;
  \item for $\gamma \in \mathbb R$ we can compute the \emph{scalar product} $\gamma \CC$ with
        \[
           (\gamma \CC)[\bar y] := \gamma \CC[\bar y] \quad \forall \bar y \in V(\GG)^r
        \]
        in time $O(r |\CC|)$.
\end{itemize}

\noindent
Given two counting data structures  $\CC_1, \CC_2$ of depth $\geq r$ the following
two operations must be supported:
\begin{itemize}
  \item The \emph{$r$-depth difference} $\CC_1 -_r \CC_2$ with
        \[
          (\CC_1 -_r \CC_2)[\bar y] := \CC_1[\bar y] - \CC_2[\bar y] \quad \forall \bar y \in V(\GG)^r
        \]
        in time $O(r \cdot \max(|\CC_1|,|\CC_2|))$;
  \item the \emph{$r$-depth product}
        $\CC_1 \conv_r \CC_2$ with
        \[
          (\CC_1 \conv_r \CC_2)[\bar y] := \CC_1[\bar y] \cdot \CC_2[\bar y] \quad \forall \bar y \in V(\GG)^r
        \]
        in time $O(r \cdot \max(|\CC_1|,|\CC_2|))$.
\end{itemize}

\noindent
A convenient way to implement $\CC$ is a prefix-trie in which every node
contains a counter (which contains the sum-total of all values stored below
it) and a dynamically sized hash-map to store its descendants. It is trivial
to update the counters during an increment and answering the prefix query
$\CC[\bar y]$ amounts to locating the node with prefix $\bar y$ in $\CC$ and
returning its counter in time $O(r)$.

Since we can easily enumerate all keys contained in $\CC$ by a depth-first
traversal, implementing the scalar product can be done by first creating an
empty counting data structure $\CC'$ and inserting all keys~$\bar x$ contained
$\CC$ by incrementing the value of $\CC'[\bar x]$ by $\gamma \CC[\bar x]$. The
DFS on $\CC$ takes time $O(|\CC|)$ and each insertion takes time $O(r)$,
hence the claimed running time holds true.

To perform the $r$-depth difference and product we traverse the two tries
$\CC_1$ and $\CC_2$ in lockstep, meaning that we only descend in the DFS if
the two currently active nodes $x_1$ in $\CC_1$ and $x_2$ in $\CC_2$ both have
a child with the same respective key, and truncating the DFS at depth $r$. During
this traversal, it is easy to populate a new trie to obtain the final result
$(\CC_1 -_r \CC_2)$ or $(\CC_1 \conv_r \CC_2)$. The lockstep DFS takes times
$O(\max(|\CC_1|,|\CC_2|))$ each insertion into the resultant trie takes time
$O(r)$ and the running time follows.

\begin{figure}[htb]
  \centering
  \includegraphics[scale=.65]{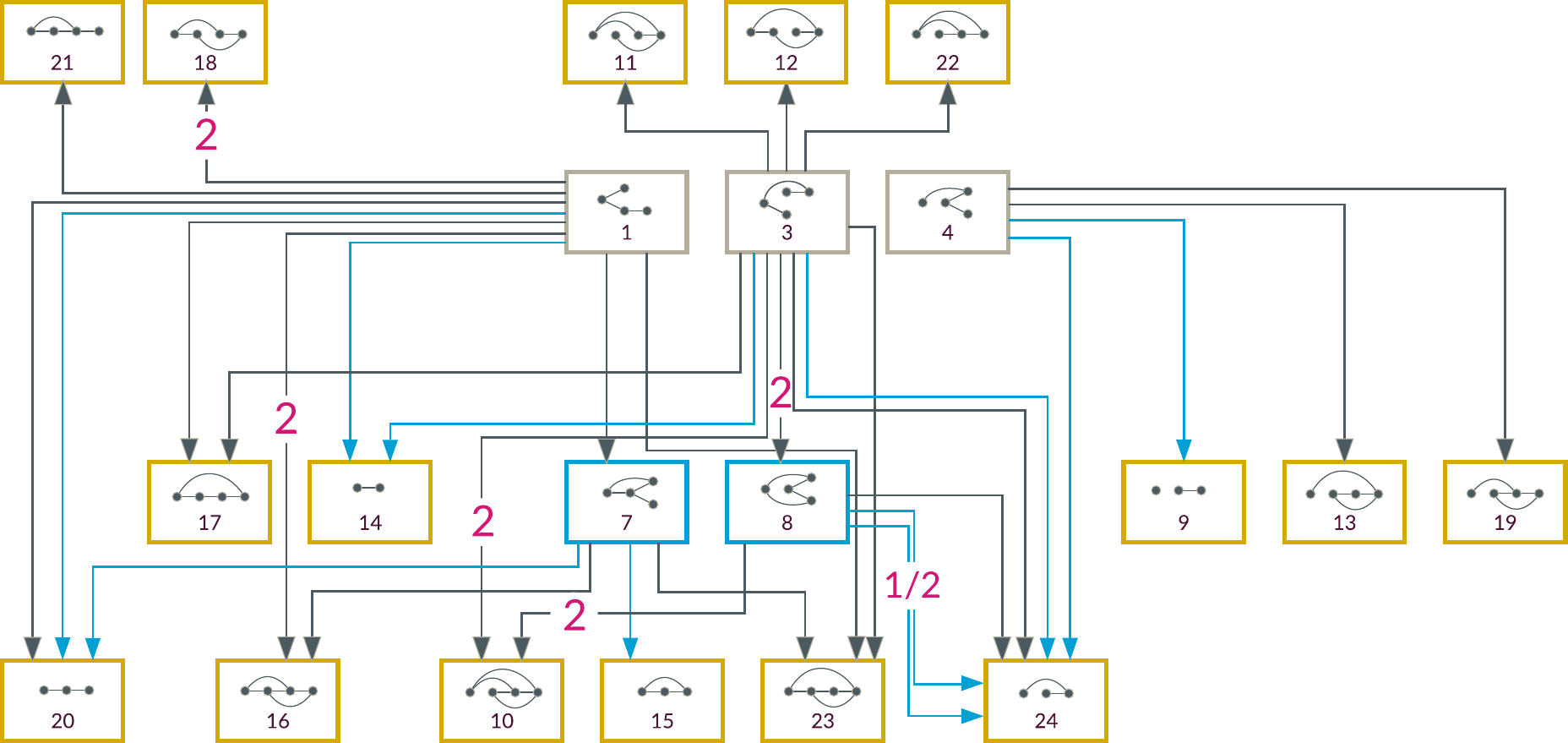}
  \caption{\label{fig:dag}%
  Task-dag for counting a path on four vertices, as depicted in Figure~\ref{fig:P4}. Blue arcs belong to edges~$E^\times$, gray arcs to $E^-$. Note that the coefficients are for the automorphism-corrected counts, therefore the hyperedges $E^\times$ need to be imbued with a weight as
  well.
  }
\end{figure}

\subsection*{The counting dag}

A \emph{counting dag} is a directed hypergraph $(\mathcal V, E^\times, E^-)$ with
two types of edges. $E^\times \subseteq \mathcal V^3$ contains edges of the
form $(\H,\H_l,\H_r)$ with $\H = \H_r \oplus_{\bar x} \H_l$ which indicate
that in order to compute $\hash_{\bar x \mapsto \any}(\H)$ by application of
Lemma~\ref{lemma:counting}, we need to compute $\hash_{\bar x \mapsto
\any}(\H_l)$ and $\hash_{\bar x \mapsto  \any}(\H_r)$ first because they appear
in the product on the right hand side. Every node in $\mathcal V$ has at
most one outgoing hyperedge in $E^\times$. Also note that $\H_l = \H_r$ is
possible.

\begin{algorithm}[ht]
  \DontPrintSemicolon
  \varrule[\heavyrulewidth] \\
  \KwIn{An etog $\H$.}
  \KwOut{A counting dag $\dir C(\H)$.} \vspace*{-6pt}
  \varrule[\lightrulewidth] \\

  \SetKwFunction{decompose}{decompose}
  \SetKwProg{Fn}{function}{}{}
  \Fn{\decompose{$\H$}}{
    Initialize $\dir C$ as an empty counting dag\;
    Let $\bar x = \stem(\H)$ \;
    Choose decomposition $\H = \H_1 \oplus_{\bar x} \H_2$\;

    $\dir C \leftarrow \dir C \cup \decompose(\H_1)$\;
    $\dir C \leftarrow \dir C \cup \decompose(\H_2)$\;
    $E^\times(\dir C) \leftarrow E^\times(\dir C) \cup \{(\H,\H_1,\H_2)\}$\;

    \For{$\D \in \defects(\H_1,\H_2)$}{
      $\eta \leftarrow \hash_{\bar x \mapsto \bar x}(\H,\D\mid \H_1,\H_2)$\;
      $\alpha \leftarrow \hash_{\bar x \mapsto \bar x}(\D,\D)$\;
      $\dir C \leftarrow \dir C \cup \decompose(\D)$\;
      $E^-(\dir C) \leftarrow E^-(\dir C) \cup \{(\H,\D,\eta/\alpha^2)\}$\;
    }
    \Return $\dir C$\;
  }

  \varrule[\heavyrulewidth] \\[1ex]
  \caption{\label{alg:decompose}%
    Recursive computation of a counting dag.
  }
\end{algorithm}

Similarly, $E^- \subseteq \mathcal V^2 \times \mathbb R$ contains
edges of the form $(\H, \D, \gamma)$ where $\D$ is a defect of $\H_l, \H_r$
and in order to compute $\hash_{\bar x \mapsto \any}(\H)$ from
$\hash_{\bar x \mapsto \any}(\H_l) \cdot \hash_{\bar x \mapsto \any}(\H_r)$
we need to subtract $\gamma \cdot \hash_{\bar x \mapsto \any}(\D)$ for all
such edges.

For two counting dags $\dir C, \dir C'$ we write $\dir C \cup \dir{C'}$ to denote the union of their vertex and edge sets. With that notation in
mind, Algorithm~\ref{alg:decompose} shows how to compute a counting dag for a
given etog $\H$. Note that the choice of decomposition $\H = \H_1 \oplus_{\bar
x} \H_2$ is arbitrary, reasonable choices include either letting $\H_1$ be as
small as possible or trying to balance the size of $\H_1$ and $\H_2$.

\begin{lemma}\label{lemma:compute-counting-dag}
  Given a graph~$H$, we can construct a counting dag
  $\dir C(H)$ with $\|\dir C\| \leq 4^{h^2}$  using Algorithm~\ref{alg:decompose} in time~$O(\|\dir C\|)$.
\end{lemma}
\begin{proof}
  We enumerate the at most~$h!$ etogs of~$H$ and run
  Algorithm~\ref{alg:decompose}, then we take the union of all resulting counting
  dags to obtain~$\dir C := \dir C(H)$. If we employ memoization across the
  calls, the total running time is bounded by~$\|\dir C\| \leq 4^{h^2}$.
\end{proof}

\subsection*{The algorithms}

\begin{algorithm}[tbh]
  \DontPrintSemicolon
  \varrule[\heavyrulewidth] \\
  \KwIn{A linear host graph $\GG$ and a counting dag~$\dir C(H)$ of a pattern~$H$.}
  \KwOut{The number of embeddings of $H$ into induced subgraphs of $G$}
  \varrule[\lightrulewidth] \\
  \BlankLine

  \SetKwFunction{decompose}{decompose}

  \comment{1}{Initialize counting data structures} \\
  Compute topological ordering $\H_1,\ldots,\H_\ell$ of $V(\dir C)$ such that $\H_1,\ldots,\H_s$ are source-nodes in $\dir C$ and
  $\H_t,\ldots,\H_\ell$ are sink-nodes in $\dir C$\;
  Initialize counting data structures $\CC_i$ of depth $|\stem(\H_i)|$ for $i \in [1,\ell]$\;

  \comment{2}{Count linear patterns} \\
  \For{$i \in [t, \ell]$}{
    \For{$v \in \mathcal \GG$}{
      \comment{3}{Count patterns ending in $v$ using weak reachability} \\
      $W \leftarrow W^{|\H|}_\GG(v)$ \;
      \For{$\bar y \in W^{|\H_i|-1}$}{
        \If{$\H_i \embeds{V(\H) \mapsto \bar yv} \GG$}{
          $\CC_i[\bar yv] \leftarrow \CC_i[\bar yv] + 1$;
        }
      }
    }
  }

  \comment{4}{Propagate counts} \\
  \For{$i \in (t-1,t-2,\ldots,1)$}{
    Let $l,r$ be the indices for which $(\H_i,\H_l,\H_r) \in E^\times(\dir C)$\;
    $k \leftarrow |\stem(\H_i)|$\;
    $\CC_i \leftarrow \CC_l \,\conv_k\, \CC_r$\;
    \For{$(\H,\H_d,\gamma) \in E^-(\dir C)$}{
      $\CC_i \leftarrow \CC_i -_k \gamma \CC_d$\;
    }
  }

  \comment{5}{Sum up counts in sink-nodes} \\
  $c \leftarrow 0$\;
  \For{$i \in (s,\ldots,1)$}{
    $c \leftarrow \CC_i[\emptyset]$\;
  }
  \Return c\;

  \varrule[\heavyrulewidth] \\[1ex]
  \caption{\label{alg:count}%
    The subgraph counting algorithm using weak reachability. Note that part
    \protect \commentmark{1} is independent of $\GG$, hence
    the counting dag $\dir C$ for any given pattern graph $H$ can
    be precomputed. }
\end{algorithm}

\begin{algorithm}[tbh]
  \DontPrintSemicolon
  \varrule[\heavyrulewidth] \\
  \KwIn{A linear host graph $\GG$ and a counting dag~$\dir C(H)$ of a pattern~$H$.}
  \KwOut{The number of embeddings of $H$ into induced subgraphs of $G$}
  \varrule[\lightrulewidth] \\
  \BlankLine

  ~~$\vdots$\\[4pt]

  \comment{2}{Count linear patterns} \\
  \For{$i \in [t, \ell]$}{
    \For{$v \in \mathcal \GG$}{
      \comment{3}{Count patterns ending in $v$ using strong reachability} \\
      $h \leftarrow |\H_i|$\;
      \For{$x_{h-1} \in S_\GG^{|H|}[v]$}{
        $j \leftarrow \hint^\H_{\H_i}(h-2)$ \;
        \For{$x_{h-2} \in S_\GG^{|H|}[x_j]$}{
          $\ddots$ \\
          $j \leftarrow \hint^\H_{\H_i}(2)$ \;
          \For{$x_1 \in S_\GG^{|H|}[x_j]$}{
              \If{$\H_i \isom \GG[x_1,\ldots,x_p]$}{
                $\CC_i[x_1,\ldots,x_p] \leftarrow \CC_i[x_1,\ldots,x_p] + 1$\;
              }
          }
          \reflectbox{$\ddots$} \\
        }
      }
    }
  }

  ~~$\vdots$\\
  \varrule[\heavyrulewidth] \\[1ex]
  \caption{\label{alg:count-col}%
    Modification of Algorithm~\ref{alg:count} to use strong instead
    of weak reachability. For ease of presentation, the algorithm is shown
    as a sequence of nested loops instead of recursion or a loop with a stack
    of partial solutions.
  }
\end{algorithm}

\noindent
The following proofs of the worst-case running time are not very indicative of
the algorithms performance as a) the term~$4^{h^2}$ is a (crude) upper bound
on the size of the counting dag and b) not every pattern of size~$h$ needs to
use the $(h-1)$-reachable sets. As an extreme example, the graph~$K_h$ only
needs $1$-reachable sets, \eg only a degeneracy ordering of the host graph.
We include a table with sizes of counting dags and the necessary depth for various patterns graph in the subsequent section.

\begin{lemma}
  Algorithm~\ref{alg:count} computes the number of induced embeddings of $H$
  into $G$ in time $O(\|\dir C\| \cdot h  \wcolnum_{h}(\GG)^{h-1} |G|)
  = O(4^{h^2} \cdot h  \wcolnum_{h}(\GG)^{h-1} |G|)$ where $h
  := |H|$.
\end{lemma}
\begin{proof}
  In part \commentmark{1}, for
  every one of the $\ell := \leaves(\dir C)$ many sinks~$\H_i$, $i \in [t,\ell]$,
  of $\dir C$ we fill the counting data structure $\C_i$ in time
  $O(\wcolnum_{h}(\GG)^{h-1} |G|)$ by application of
  Lemma~\ref{lemma:linear-count-wcol}.

  Since every counting data structure at the end of part \commentmark{2}
  contains at most $O(\wcolnum_{h}(\GG)^{h-1} |G|)$ many tuples, it follows
  that all operations in step \commentmark{4} on counting data structures
  ($r$-depth products, differences, scalar products) can be computed in time
  $O(h \wcolnum_{h}(\GG)^{h-1} |G|)$. The number of such operations
  is proportional to~$\|\dir C\|$, thus in total step \commentmark{4}
  takes times
  $
    O(\|\dir C\| \cdot h  \wcolnum_{h}(\GG)^{h-1} |G|)
  $. The time taken in step $\commentmark{5}$ is negligible compared to the
  previous steps and we conclude that the total running time is as claimed.

  The correctness of the algorithm follows by induction over the counting
  dag: the leaf counts are correct by Lemma~\ref{lemma:linear-count-wcol}
  and the counts at the internal nodes are correct by Lemma~\ref{lemma:counting}.
\end{proof}

\noindent
Combining the above lemma with Propositions~\ref{prop:col-adm}
and~\ref{prop:adm-linear-time}, we immediately obtain the following:

\begin{corollary}
  Let~$\mathcal G$ be a bounded expansion class. There exists an algorithm that
  for every graph~$H$ on $h$ vertices and~$G \in \mathcal G$, computes the
  number of times~$H$ appears as an induced subgraph in~$G$ in total time
  $O(4^{h^2}h (\wcolnum_h(G)+1)^{h^3} |G|)$.
\end{corollary}

\noindent
Exchanging Lemma~\ref{lemma:linear-count-wcol} for
Lemma~\ref{lemma:linear-count-col} in the above proof shows a similar running
time for the variants using strong reachability:

\begin{lemma}
  Algorithm~\ref{alg:count-col} computes the number of induced embeddings of $H$
  into $G$ in time $O(\|\dir C\| \cdot h  \colnum_{h}(\GG)^{h-1} |G|)
  = O(4^{h^2} \cdot h  \colnum_{h}(\GG)^{h-1} |G|)$ where $h
  := |H|$.
\end{lemma}

\begin{corollary}
  Let~$\mathcal G$ be a bounded expansion class. There exists an algorithm that
  for every graph~$H$ on $h$ vertices and~$G \in \mathcal G$, computes the
  number of times~$H$ appears as an induced subgraph in~$G$ in total time
  $O(4^{h^2}h \colnum_h(G)^{h^2} |G|)$.
\end{corollary}

\section{Discussion}\label{sec:discussion}

We begin by discussing our proof-of-concept implementation\footnote{Code
available under a BSD 3-clause license at
\url{http://www.github.com/theoryinpractice/mandoline}.} along with preliminary
experimental results. Several observations on natural extensions of this
algorithm follow.

In practice, Algorithm~\ref{alg:count} and~\ref{alg:count-col} have a lot of
engineering potential. In most cases, the search space for linear patterns is
much smaller than the $h$-weak or strong neighbourhoods since previously-fixed
vertices will often have the sought vertices in their left neighbourhood or in
a weak/strong neighbourhood at distance less than~$h$.
Furthermore, the task dag for a given pattern can be precomputed and optimized;
in order minimize memory use, we can process tasks in an order which enables us to delete counting
data structures as soon as they have been propagated along all in-edges.

Since we view the counting dag computation as a form of pre-processing,
we implemented this stage using Python and show results for
several small pattern graphs in Table~\ref{tab:taskdag}. We have not yet explored
whether different decomposition strategies (i.e. which piece-sum decomposition to
choose if there are multiple options) significantly impact the size of these
dags. As expected, the counting dag is smaller for denser graphs---for complete
graphs the algorithm essentially reduces to the well-known clique-counting
algorithm for degenerate graphs.

\begin{table}[hbt]
   \centering
  \begin{tabular}{llll}
    $G$ & $|\dir C|$ (leaves) & $\|\dir C\|$ & $d$ \\ \midrule
    $P_{3}$ & 5 (4) & 3 & 1 \\
    $P_{4}$ & 25 (20) & 26 & 2 \\
    $P_{5}$ & 247 (186) & 552 & 3 \\ \midrule
    $C_{3}$   & 1 (1) & 0 & 1 \\
    $C_{4}$ & 5 (4) & 3 & 1 \\
    $C_{5}$ & 32 (27) & 27 & 2 \\
    $C_{6}$ & 424 (338) & 689 & 2 \\ \midrule
    $S_{3}$ & 14 (9) & 21 & 1 \\
    $S_{4}$ & 60 (36) & 200 & 2 \\
    $S_{5}$ & 619 (389) & 4919 & 2 \\ \midrule
    $W_{3}$ & 1 (1) & 0 & 1 \\
    $W_{4}$ & 21 (18) & 9 & 1 \\
    $W_{5}$ & 141 (123) & 90 & 2 \\
    $W_{6}$ & 1707 (1395) & 2332 & 2 \\ \midrule
    $K_i$   & 1 (1) & 0 & 1 \\ \midrule
    $K_{2,2}$ & 5 (4) & 3 & 1 \\
    $K_{3,3}$ & 24 (17) & 27 & 1 \\
    $K_{4,4}$ & 132 (87) & 281 & 2 \\
    $K_{5,5}$ & 890 (620) & 1570 & 2 \\
  \end{tabular}\hspace*{.4cm}%
  \begin{tabular}{lllll}
    $n$ &$G$ & $|\dir C|$ (leaves) & $\|\dir C\|$ & $d$ \\ \midrule
    4 & diamond & 8 (7) & 3 & 1 \\
     & paw & 18 (15) & 12 & 1 \\ \midrule
    5 & butterfly & 56 (44) & 85 & 2 \\
     & gem & 90 (77) & 61 & 2 \\
     & cricket & 94 (65) & 226 & 2 \\
     & house & 110 (92) & 88 & 2 \\
     & dart & 121 (93) & 171 & 2 \\
     & kite & 141 (116) & 175 & 2 \\
     & bull & 199 (154) & 325 & 2 \\ \midrule
    6 & co-net & 371 (306) & 441 & 2 \\
     & domino & 723 (572) & 1110 & 2 \\
     & co-domino & 733 (606) & 1050 & 2 \\
     & co-fish & 908 (734) & 1515 & 2 \\
     & net & 1805 (1388) & 4333 & 3 \\
     & fish & 2052 (1556) & 5436 & 3
    \vspace*{2.2cm}
  \end{tabular}
  \caption{\label{tab:taskdag}%
    Size and number of leaves for counting dags for various small graphs. The
    final column gives the reachability-depth~$d$ necessary to count the specified pattern.
    Named graphs can be looked up under \protect\url{http://www.graphclasses.org/smallgraphs.html}.
  }
\end{table}

\noindent
For the subgraph counting algorithm, we chose to implement in Rust. While we recognize
that there is significant additional optimization and engineering needed, it is
notable that runtimes remain reasonable (see Table~\ref{tab:runtime}) on host
graphs with tens of thousands of nodes even for relatively large
patterns (all measurements where taken on a simple laptop with an intel i5 core
and 4GB RAM).

\begin{table}
\begin{tabular}{lllllll}
  Network         &  $n$    &  $m$    &  $P_5$  &  $W_5$ &  bull  &  $K_{4,4}$ \\  \midrule
  soc-advogato    &  6551   &  43427  &  4m36s  &  1m17s &  2m23s &  1m5s   \\
  cora-citation   &  23166  &  89157  &  3m23s  &  2m8s  &  2m58s &  1m33s  \\
  ca-CondMat      &  23133  &  93497  &  3m26s  &  1m57s &  2m41s &  1m21s  \\
  Google+         &  23628  &  39194  &  2m57s  &  1m45s &  2m13s &  1m18s  \\
  digg            &  30398  &  86312  &  5m50s  &  2m53s &  3m38s &  2m8s
\end{tabular}
\caption{\label{tab:runtime}%
  Runtimes for counting several common patterns in five real-world networks.
}
\end{table}

We plan to engineer these implementations further and compare it to other
subgraph-counting algorithms on a larger corpus of host and pattern graphs
in future work.
Note that it is straightforward to extend our algorithm to edge- and
vertex-labelled graphs by defining isomorphisms and embeddings
appropriately. We chose not to include labels here
as they add another layer of notation that would make the presentation less clear.

\noindent 
We also, for simplicity, assumed that the pattern graph $H$ is connected. This
is easily remedied by a labelled version of the algorithm: we add an apex
vertex with a unique label to both $H$ and $G$ and make it the minimum in
$\GG$. Alternatively, the presented algorithm can be modified by allowing
piece-sums to work on connected components. This modification does not significantly
change the algorithm, but adds additional cases in many proofs.

Finally, we observe that the approach presented here can be modified to count non-induced
subgraphs, subgraph homomorphisms or boolean queries instead by adjusting the
notions of patterns and pattern decompositions appropriately.

\bibliographystyle{plain}
\bibliography{biblio}

\begin{thebibliography}{10}

\bibitem{SGC}
C.~T. Brown, D.~Moritz, M.~P. O'brien, F.~Reidl, T.~Reiter, and B.~D. Sullivan.
\newblock Exploring neighborhoods in large metagenome assembly graphs reveals
  hidden sequence diversity.
\newblock {\em BioRxiv}, page 462788, 2019.

\bibitem{CountingPathHard}
Y.~Chen and J.~Flum.
\newblock On parameterized path and chordless path problems.
\newblock In {\em Twenty-Second Annual IEEE Conference on Computational
  Complexity (CCC'07)}, pages 250--263. IEEE, 2007.

\bibitem{SparsityNetworks}
E.~D. Demaine, F.~Reidl, P.~Rossmanith, F.~{S{\'a}nchez Villaamil}, S.~Sikdar,
  and B.~D. Sullivan.
\newblock Structural sparsity of complex networks: Bounded expansion in random
  models and real-world graphs.
\newblock {\em Journal of Computer and System Sciences}, 2019.

\bibitem{DvorakDomset}
Z.~\Dvorak.
\newblock Constant-factor approximation of the domination number in sparse
  graphs.
\newblock {\em European Journal of Combinatorics}, 34(5):833--840, 2013.

\bibitem{SubgraphDynamic}
Z.~\Dvorak and V.~\Tuma.
\newblock A dynamic data structure for counting subgraphs in sparse graphs.
\newblock In {\em Workshop on Algorithms and Data Structures}, pages 304--315.
  Springer, 2013.

\bibitem{EppsteinDegenCliques}
D.~Eppstein, M.~L{\"{o}}ffler, and D.~Strash.
\newblock Listing all maximal cliques in sparse graphs in near-optimal time.
\newblock In {\em International Symposium on Algorithms and Computation}, pages
  403--414. Springer, 2010.

\bibitem{ParamCounting}
J.~Flum and M.~Grohe.
\newblock The parameterized complexity of counting problems.
\newblock {\em SIAM Journal on Computing}, 33(4):892--922, 2004.

\bibitem{FlumGrohe}
J.~Flum and M.~Grohe.
\newblock {\em Parameterized Complexity Theory (Texts in Theoretical Computer
  Science. An EATCS Series)}.
\newblock Springer, 2006.

\bibitem{FOEnumBndExp}
W.~Kazana and L.~Segoufin.
\newblock Enumeration of first-order queries on classes of structures with
  bounded expansion.
\newblock In {\em Proceedings of the 32nd {ACM} {SIGMOD-SIGACT-SIGART}
  Symposium on Principles of Database Systems (PODS)}, pages 297--308. {ACM},
  2013.

\bibitem{WCol}
H.~A. Kierstead and D.~Yang.
\newblock Orderings on graphs and game coloring number.
\newblock {\em Order}, 20(3):255--264, 2003.

\bibitem{WcolExperimental}
W.~Nadara, M.~Pilipczuk, R.~Rabinovich, F.~Reidl, and S.~Siebertz.
\newblock Empirical evaluation of approximation algorithms for generalized
  graph coloring and uniform quasi-wideness.
\newblock 103:14:1--14:16, 2018.

\bibitem{Sparsity}
J.~\Nesetril and P.~{Ossona de Mendez}.
\newblock {\em Sparsity: Graphs, Structures, and Algorithms}, volume~28 of {\em
  Algorithms and Combinatorics}.
\newblock Springer, 2012.

\bibitem{TooManyColours}
M.~P. O’Brien and B.~D. Sullivan.
\newblock Experimental evaluation of counting subgraph isomorphisms in classes
  of bounded expansion.
\newblock {\em CoRR, abs/1712.06690}, 2017.

\bibitem{FelixThesis}
F.~Reidl.
\newblock {\em {S}tructural sparseness and complex networks}.
\newblock Dr., Aachen, Techn. Hochsch., Aachen, 2016.
\newblock Aachen, Techn. Hochsch., Diss., 2015.

\bibitem{WColBndExp}
X.~Zhu.
\newblock Colouring graphs with bounded generalized colouring number.
\newblock {\em Discrete Mathematics}, 309(18):5562--5568, 2009.

\end{thebibliography}

\end{document}